\documentclass{amsart}
\usepackage{amssymb}
\usepackage{latexsym}
\input xypic


\def\Z{\mathbb{Z}}

\def\C{\mathbb{C}}

\def\R{\mathbb{R}}
\def\C{\mathbb{C}}

\def\Pi{\mathbb{P}^{\infty}}

\def\Zpk{\mathbb{Z}/p^{k}}
\def\Zpk1{\mathbb{Z}/p^{k-1}}

\newcommand{\rref}[1]{(\ref{#1})}

\newcommand{\cform}[3]{\begin{array}{c}
{\scriptstyle #3}\\
#1\\
{\scriptstyle #2}\end{array}}

\newcommand{\fracd}[2]{\frac{\displaystyle #1}{\displaystyle #2}}

\newcommand{\beg}[2]{\begin{equation}\label{#1}#2\end{equation}}
\def\r{\rightarrow}

\def\mh{\mathcal{H}}

\def\sl2{\widetilde{SL_{2}(\Z)}}

\begin{document}
\title[The Kondo effect WZW brane formalism]{A mathematical formalism for the
Kondo effect in WZW branes}
\author{Po Hu and Igor Kriz}\thanks{The authors were supported by the NSF}

\maketitle

\section{Introduction}

\label{s1}

The aim of this paper is to give a setting
which would allow a mathematically rigorous
treatment of the theory of WZW $D$-branes.
In particular, we apply (with some changes) the formalism developed in
\cite{hk} to capturing the WZW $D$-brane picture.
The theory of WZW branes has several components 
and has been previously worked
out quite satisfactorily physically (see e.g. 
\cite{cardy,al1,al2,sch,fs,mms,moore,fffs,stan,bds,rid}). The physical answer is that stable
$D$-branes (at least on CFT level)
in the level $k$ WZW model corresponding to a group $G$ are classified by 
a suitable twisted $K$-theory group\footnote{The groups \rref{etw1}
are now completely known for $G$ a simple Lie group: V. Braun 
\cite{braun}, using an idea of Schafer-Nameki \cite{sakura}, calculated
the answer with the assumption that the Verlinde algera is a complete
intersection ring. A proof removing that assumption
was independently obtained by Chris Douglas \cite{douglas}.}
\beg{etw1}{K_{\tau}(G).} 
However, as is to be expected, it is not trivial to 
interpret this classification result mathematically. In fact, the present
paper was motivated by a question of Mike Hopkins whether there is an
analogous geometric interpretation of the twisted $K$-group $K_{\tau}(G)$
as the previous complete calculation of the equivariant twisted $K$-group
$K_{\tau}^{G}(G)$ as the Verlinde algebra by Hopkins, Freed and
Teleman \cite{hft} (the twisting $\tau$ corresponds
to the level or the level shifted by the dual Coxeter number, depending whether or
not supersymmetry is involved, respectively). Although physically the answer
to this question is known as to be given by CFT-stable WZW $D$-branes,
for the purposes of a mathematician, the answer must be revisited and 
stated in precise mathematical terms.

\vspace{3mm}
The goal of the present paper is to advance this program
by stating at least a precise mathematical conjecture: We would like to conjecture
that the group of charges of WZW branes is equal to the twisted
$K$-group proposed, but at present there is no precise mathematical definition
of one side of the equation, namely the group of charges of $D$-branes
(other than, of course, a definition which would input the answer desired).
In the previous paper \cite{hk}, the authors proposed a mathematical
definition of $D$-brane charges, but this definition does not capture
an important physical aspect of the WZW situation, namely the Kondo
flow of Affleck and Ludwig \cite{al1,al2}. These flows mean that
we must allow, as a part of the formalism, 
transition theories which break conformal invariance
on the boundary. Without such flows (or some other additional
input, such as spacetime symmetries), the physical answer would not be the 
non-equivariant twisted
$K$-group conjectured.

\vspace{3mm}
In the present paper, then, we develop a variant of the formalism
of \cite{hk} which allows an axiomatization of the flows. Therefore,
we present here a setting in which the ``group of $D$-brane charges'' is 
well defined. We do not prove in this paper that the group of charges
is equal to the twisted $K$-group. Thus, our main result is formulating
the conjecture. However, there is one concrete piece of the puzzle which
we do fill in: even physically, currently little is known about the renormalization
of the Kondo flows. In other words, not much is known about the nature
of the flow. It is, for example, not known that the theories with broken
boundary conformal invariance which arise along the flow are well defined
field theories with convergent amplitudes. This question, in and of
itself, is hard, and will also remain conjectural in the present paper.
However, as evidence of the conjecture, we will construct 
rigorously mathematically one element of the conjectured structure of
the theories along the flow, namely the ``deforming field'' at each point
of the flow. Because we do not construct the whole deformed theory, the precise
setting in which this field will be constructed will have to be specified later.
We should mention, however, that this does provide a firm starting point
for a mathematical
foundation of calculations, as with knowing the deforming field, 
vacua of the deformed theory are obtained as solutions of differential
equations, and hence our result essentially puts the Kondo flow on
equal footing with a renormalized perturbative quantum field theory
(although the formalism is somewhat different). 

\vspace{3mm}
To be even more concrete
about what is to be done, we must understand the elements of the 
WZW model in more detail.
First, there is the question of a mathematical
formalism for conformal field theory (CFT). In the present paper,
we use the formalism proposed by G. Segal \cite{scft}, and written 
down in detail in \cite{hk,hk1,fi},
using our formalism of SLCMC's (explained below in Section \ref{s1a}).
(Actually, for a CFT with $1$-dimensional anomaly, the formalism of \cite{scft}
is sufficient; subtleties arise in the case of chiral CFT, which will come
into play later.) The reason we choose the Segal formalism is that
it is in some sense the most ``maximalistic'', i.e. it models
most of the structure desired, and the axioms can be traced most naively,
i.e. with least input, to fundamental physical principles. The reason
the work of \cite{hk,hk1,fi} became necessary is that if one wants to
capture rational conformal field theory and modular functors (which is
relevant here), the notion of ``equality'' of some of the spaces involved
becomes delicate, and leads to $2$-category theory. In fact, Fiore \cite{fi1}
recently showed that even a simplified ``cobordism'' approach to CFT proposed
in \cite{scft} leads to $2$-category theory when completely
rigorized on this level. The axiomatizations of \cite{hk,hk1,fi,fi1}
proceed in the direction of the same philosophy, i.e. the axiomatization
is ``zero input'', it uses only Segal's geometric principles, and
generic $2$-category formalism. Since developing this theory, the
authors discovered that an alternate rigorous approach to modular functors
was previously developed by Bakalov-Kirillov \cite{bk} following Turaev
\cite{tur}. These formalisms, however, use more input in the sense that
more sophisticated concrete axioms have to be known and
stated upfront. 

\vspace{3mm}

A major alternative of Segal's axioms are the axioms of vertex operator
algebra, which is originally designed to model a part of the structure
of the genus $0$ part of a chiral CFT. Actually the axioms (and also
terminology) vary somewhat, but the basic mathematical references 
are \cite{flm,bor}. Recently \cite{hk}, this formalism has been adapted
to work for the genus $0$ part of a conformal field theory
with both chiralities. Extensions of information to higher genus
has also been discussed in \cite{gg,ty,huang}. Other references include
\cite{fbz,hkong1,longreh} and many others. These works include discussions
of boundary sectors, although we are not aware of any references
discussing flows in this
axiomatic setting. An important point is that all of these approaches
set out to capture only parts of the ``naive'' structure one sees
in the Segal approach. With all the recent advances mentioned, it
is quite possible that in a suitable setting, both approaches can
be proved to be equivalent, but as far as we are aware, no such
theorems are known at present. On the other hand, it is also important
to mention that the VOA approach captures certain situations
the Segal approach cannot, for basic substantive reasons. For example,
the Segal approach, following the idea that CFT should be a ``quantum
mechanics of strings'', uses state spaces which are Hilbert spaces, but
if one wants to discuss CFT on spacetime with indefinite signature,
and BRST, one definitely cannot use Hilbert spaces, and the ``naive''
conformal invariance of the Segal approach is too much to expect. 
This is the mathematical explanation of the one-loop divergence
of bosonic string theory. On the other hand, these questions
can be certainly discussed in the vertex operator formalism.
Additionally, the vertex operator formalism is generally much more successful
for the purpose of concrete calculations. It is also not known
whether a conformal field theory in the sense of Segal always gives
rise to a vertex operator algebra: this requires taking limits
of ``distributional type'', whose existence is difficult to prove without
additional assumptions.
In this paper, we will also resort to the vertex operator formalism
at one point, namely when constructing the deforming field of the
Kondo flow. Proving that the flow exists in the Hilbert space setting
would require additional work which at the moment is conjectural.

\vspace{3mm}
Next thing to discuss is 
the WZW model, which will be described in section \ref{s2a}, 
and which is to be our main example. As a physical
theory, this CFT has been constructed in \cite{wzw}. 
There have been many developments since. Mathematically, the
corresponding vertex operator algebra has been constructed
by Frenkel and Zhu \cite{fzhu}, and the conformal
blocks were constructed by Tsuchiya-Ueno-Yamada
\cite{tuy}. Physically, the spectrum of primary
fields was derived by Felder-Gawedzki-Kupiainen \cite{fgk}.
Another important contribution is the free field
realization of the WZW model, which is,
including the non-zero labels, is described well in 
\cite{gh}. 
The modular functor of the WZW (in a sense
equivalent to ours) have been formalized in \cite{bk}.
Although a 
formal proof that the WZW model satisfies Segal's
axioms, as far as we know, has never been published, 
we believe that a proof can be pieced together
from the sources cited, since the main point
particular to the Segal approach is convergence of
the vacua in the Hilbert space setting,
which actually follows from the Coulomb gas realization
(in fact, it shows that the vacua are ``smooth''
elements, which is a much stronger statement
than Hilbert space convergence). We will return to
this point in Section \ref{s2a} below. Nevertheless, 
checking all the details of the comparisons
of the relevant setups in all of the points needed to show
that the WZW model is a Segal CFT is a
subtle proposition, which we feel is somewhat independent
of the main subject of this paper, which is a mathematical
axiomatization of the flows. For this reason, the discussion of the
Segal axioms for the WZW in this paper stops short of a mathematical theorem,
and technically, the statement remains a conjecture. We believe
that the remaining steps are routine, and, additionally, really
belong to other authors. On the other hand, we do describe, in
any RCFT, a construction of open sector Ishibashi and Cardy vacua,
by adapting the constructions of \cite{bcdcd,fffs} to the worldsheets
considered in the Segal approach. In the context of general RCFT,
Hilbert space
convergence of these vacua follows from the convergence
of the closed chiral vacua, which
again in the case of the WZW
follows from the Coulomb gas realization. Therefore, having
constructed the Hilbert vacua, together with
the published results \cite{fffs} on open/closed modular functors, we have
all the ingredients for proving that the Cardy branes form 
a consistent D-brane category in our precise mathematical sense.
Again, however, we do not state this as a theorem, and technically
only conjecture that the construction we describe satisfies all the
axioms stated.

\vspace{3mm}
Next, one must discuss mathematically the
axiomatization of the notion of $D$-brane. A mathematical notion
of $D$-brane category from our viewpoint, extending Segal's
axioms, has been given in \cite{hk} (there are variants, we shall explain
in Section \ref{s2} below which variant we use here and why). 
An additional complication
is that we do not have a mathematical interpretation for the question 
what {\em are} the $D$-branes of a given CFT. It is in principle
possible that even if we fix a set of $D$-branes $a$ and know each state space
of open strings beginning and ending on $a$, the state spaces
of open strings stretched between different branes could vary: in physical
language, this is expressed by the statement that branes violate locality
somewhat. Instead of conjecturing that the WZW $D$-brane charges
are classified by the twisted $K$-theory \rref{etw1}, one must
therefore really conjecture that there exists a consistent system
of $D$-branes whose charges are classified by the conjectured group.

\vspace{3mm}
To be precise, we must again distinguish whether supersymmetry (SUSY) is present.
When we work with the supersymmetric WZW model, the twisting in \rref{etw1}
is simply by $k$ times the canonical generator
of the group of twistings. This case is more interesting
physically, as one can then (at least for appropriate values
of the central charge) embed the 
WZW model into an actual critical superstring theory and consider
non-anomalous string branes. However, axiomatizing supersymmetric CFT 
and superstring theory mathematically
brings even more complications, (although we do not
feel substantial new concepts are needed); for this reason, in the present paper,
we suppress SUSY, and consider only the bosonic WZW model. The
conjectured groups of charges in the supersymmetric and
non-supersymmetric case run in parallel, with the twisting 
shifted. Thus, in the bosonic case (which we consider for reasons
of simplicity), 
the twisting $\tau$ will again be by $k+h^{\vee}$ the canonical generator
where $h^{\vee}$ is the dual Coxeter number (equal to $n$ when $G=SU(n)$).

\vspace{3mm}
To go further, we shall introduce yet another restriction: instead of
trying to interpret the entire group \rref{etw1}, we will only attempt
to interpret its subgroup corresponding to ``D0-branes''. In the case
of $G=SU(n)$, the answer should then be
\beg{etw2}{\Z/((k+n)/gcd(k+n, lcm(1,...,n-1)))
}
where $k$ is the level. The full group \rref{etw1} is \rref{etw2} tensored
with an exterior algebra, the augmentation ideal of which conjecturally
corresponds to ``higher-dimensional'' branes. In the D0-case, however,
a much more explicit physical theory is available, and so more concrete
mathematical conjectures can be stated (for example, in contrast with
the higher-dimensional case, explicit state spaces of open CFT string sectors
are known in the D0-case; in the higher dimensional case,
one encounters questions such as whether higher homology classes
are represented by submanifolds).

\vspace{3mm}
To be more precise about the sense in
which we are using the term D0-brane, as already hinted,
a part of the statement about the WZW model being a CFT
is that it is in fact a rational CFT (RCFT), i.e. is obtained by a certain
canonical procedure from a unitary chiral CFT with modular functor. Now
Cardy \cite{cardy} has proposed a general 
approach to D0-branes in RCFT. Mathematically, Cardy hasn't defined
all the structure which is required by the Segal formalism described in
Section \ref{s2}, but has defined enough of it
to make it convincing that his proposal is undoubtedly right. 
In fact, as already mentioned,
we show in Section \ref{s2a} that the
Cardy D0-branes do in fact fit our formalism by constructing
the open string vacua, so the situation is still
quite satisfactory so far. 

\vspace{3mm}
Looking at Cardy D0-branes only, however, one does not obtain the group
\rref{etw2}, rather, Cardy's D0-branes are classified by the Verlinde algebra.
To obtain the answer \rref{etw2}, one must consider {\em continuous deformation}
of D0-branes. The mathematical situation is substantially less satisfactory
here. What we can describe rigorously mathematically
(at least modulo proving convergence
of certain integrals) is {\em infinitesimal} deformation along primary
fields in the open sector. This is analogous to infinitesimal deformation
of bulk CFT as described by G.Segal in \cite{scft} (see also \cite{hk}).
We describe this construction in Section \ref{s3} below. This construction
actually has the remarkable property that when we infinitesimally deform
a brane $b$, it automatically updates also all the open string sectors, including
mixed sector between $b$ and other branes. Therefore, although we still do
not have a precise 
interpretation of the question as to what ``are'' branes,
the corresponding linearized infinitesimal question does make sense.

\vspace{3mm}
Physically,
however, infinitesimal deformations aren't enough. One is interested
in finite deformations; an intermediate step are deformations ``to perturbative
level'', which means given by Taylor expansion in the deformation parameter.
Here, however, one hits an obstruction. Namely, it turns out that the infinitesimal
boundary deformations of CFT $D$-branes cannot be always 
be continued while preserving
full conformal invariance of the boundary sectors. Therefore, we need to look
for ways to break conformal symmetry on the boundary. There is, fortunately,
a relatively easy way of doing so, by defining worldsheets where 
$D$-brane components are also parametrized like string components (with some
minor modifications). The set of such worldsheets
has the same formal properties as the set of closed/open CFT
worldsheets, and our formalism therefore allows us to define CFT with such
source worldsheets (i.e. with conformal invariance broken on the boundary).
In this more relaxed
setting, one can again ask if 
infinitesimal deformations of a brane $a$ by fields in 
the open string sector $K_{aa}$
can be ``exponentiated'', i.e. if there exist paths of such theories
where at a given point derivatives of the vacua by the path parameter
are equal to the values prescribed by the infinitesimal deformation.

\vspace{3mm} 
Affleck and Ludwig \cite{al1, al2} conjectured
that a particular such paths connects certain Cardy D-branes
in the $G$-WZW model. More specifically, for example, if we consider
an irreducible representation $V$ with weight $\lambda$, the corresponding
brane should be connected with the sum of $|V|$ copies of the brane
corresponding to the trivial irreducible representation by exponentiating,
in the latter boundary sector, the field
$$\cform{\sum}{\alpha}{}S_{\alpha}J^{\alpha}_{-1}$$
where $\alpha$ are generators of the Lie algebra $g$ corresponding
to $G$ and $S_{\alpha}$ are the corresponding elements of $Hom(V,V)$
given by the representation.
This is related to the Kondo effect modelling magnetic impurities in
superconductors and other metallic materials. 
This proposal is the basis of the answer \rref{etw2}.
It is interesting to note
that the Affleck-Ludwig proposal actually
models an effect which
can be observed in the laboratory.

\vspace{3mm}
Although we do not prove the Affleck-Ludwig conjecture here, we
work out this example in substantial detail, and actually make progress
in substantiating the process mathematically, by constructing
rigorously the whole perturbative expansion of the deformation field
of the process. To make treatment of convergence easier, we
do this calculation in the vertex operator setting and
not the Hilbert space setting. We work by induction on degree
of the perturbative parameter, and inductively prove an
estimate for the fields involved. The induction is complete
in the sense that we are able to then prove the same estimate
at the next power, so the perturbative expansion of the
deformation field is rigorously defined. We do not however prove
that the resulting vacua are represented by trace class elements in
Hilbert spaces, and do not prove that the process has the $\lambda$-brane
limit
predicted by the Kondo effect.

\vspace{3mm}
Our main point, then, is to propose 
in Section \ref{s4} an axiomatization of {\em topological
space $D$-brane category}, which is compatible with the formalism described in
Section \ref{s2}. This way, one can conjecture that there exists such category
whose set of objects has
the structure of a manifold whose tangent space is isomorphic to the space of
weight one fields considered in Section \ref{sp}, and that contains
the Cardy branes, and its set of connected components is \rref{etw2}. 

\vspace{3mm}
The present paper is organized as follows. In section \ref{s1a},
we review our formalism for axiomatizing CFT. In section \ref{s2}, we review
how this formalism applies to branes. In Section \ref{s2a}, we review Cardy branes,
boundary states, and the case of the WZW model. In Section \ref{s3}, we
describe infinitesimal deformation of branes. 
In Section \ref{sp}, we consider the Kondo effect, conformal
symmetry breaking on the boundary, and
deformations to perturbative
level. In Section \ref{s4}, we give our formalism for topological $D$-brane
category, and state a rigorous conjecture regarding
the group \rref{etw2}.

\vspace{3mm}
\noindent
{\bf Acknowledgement:} The authors are indebted to Chris Douglas and 
Sakura Schafer-Nameki for discussions on twisted equivariant $K$-theory.

\section{Preliminaries: our formalism for conformal field theory}

\label{s1a}

Our formalism for conformal field theory, rigorizing all aspects of
the definition given in \cite{scft}, can be found in \cite{hk1,hk,fi}.
The whole treatment requires a substantial amount of detail which
we do not want to reproduce in the present paper.
Because of this, we limit ourselves to an informal review. The
reader may refer to \cite{hk1,hk,fi} for a fully detailed discussion.

\vspace{3mm}
The main idea of our approach to quantum conformal field theory \cite{hk1,hk}
is that we noticed that conformal field theory, which one usually thinks
of as a fundamental object, is actually a morphism of two specimens of
the same structure:
\beg{ecft1}{\mathcal{C}\r \mathcal{V}.
}
The structure is called SLCMC (the acronym and its meaning will be explained
below). The source SLCMC $\mathcal{C}$ in \rref{ecft1} consists usually (but not
necessarily) of some type of ``surfaces with additional data'', while
the target SLCMC $\mathcal{V}$ in \rref{ecft1} consists of some type of ``state
spaces'', usually vector spaces with some additional data.

\vspace{3mm}
It should be pointed out that \rref{ecft1}
axiomatizes ``conformal field theories'' in a fairly
general sense, which exceeds the generality of the meaning
in which the term is used physically: the same formalism indeed describes
conformal field theories in the classical sense, modular functors,
conformal field theories with modular functors, closed/open conformal field theories
with various grades of anomaly allowed, etc. In other words, it is an axiomatic
setting where conformal invariance, (if it makes sense at all in the context
considered), can certainly
be broken. We shall find that capability useful later in this paper. A few words
are in order however on why use the attribute ``conformal'' at all? It may
seem that an arbitrary quantum field theory is roughly of the form \rref{ecft1}.
The main restriction on our formalism is that we wish to use the language of 
{\em stacks}. This is a categorical mechanism which allows us to introduce families of
objects, and thereby notions like continuity, analycity, holomorphy into
the formalism in a fairly easy way, without leaving the realm of algebra.
The use of stacks, if we want to use it
without further elaboration, however, requires that {\em automorphisms}
of objects of the structure under consideration (e.g. worldsheets) should be
{\em discrete}. This is typically not true when we break conformal invariance
substantially (e.g. in general QFT). On the other hand, we will see that breaking
conformal invariance on the {\em boundary} of open worldsheets only doesn't
cause such difficulties, and hence can still be captured by our formalism.
This is essential for describing the Kondo effect in WZW branes, see Sections
\ref{sp},\ref{s4} below.

\vspace{3mm}
Let us be more specific about what we mean by SLCMC, then. The acronym stands
for stack of lax commutative monoids with cancellation. As already hinted above,
the ``stack'' part is used to capture continuous families only, so let us
discuss that last. If we drop the ``stack'' attribute, we will see precisely
its sections over a point, i.e. the algebraic structure alone (e.g. the
set of all worldsheets), without 
looking at continuous families. Now such section of a stack over
a point is thought of as discrete topologically. It is not however
a set, but a category. Indeed, when looking at worldsheets which are, say,
Riemann surfaces with parametrized boundary components (as will be understood
when applying our approach to closed sector CFT), we must consider not
only the worldsheets, but their isomorphisms, which are holomorphic
diffeomorphisms compatible with boundary parametrizations.
Discrete automorphism groups occur. 

\vspace{3mm}
This is related to what ``lax'' means: when discussing any algebraic structure
in the context of categories, it is generally unreasonable to assume that
algebraic identities (such as commutativity, associativity, etc.) hold 
precisely. They generally hold only up to natural isomorphisms, called 
{\em coherence isomorphisms}, but those must in turn satisfy certain commutative
diagrams (called {\em coherence diagrams}). There is a general formalism for
how to form such diagrams, which is discussed in detail in 
\cite{hk}, \cite{hk1}, \cite{fi}. This is in general what we mean
by the word ``lax'': it means ``up to natural isomorphisms, satisfying
canonical coherence diagrams''. It should be pointed out that terminology
unfortunately
varies somewhat. In the context of category theory, what we call ``lax'' is
usually labelled by the prefixes ``pseudo-'' or ``bi-'' (see \cite{fi} for
a dictionary of terms).

\vspace{3mm}
Let us now, finally, discuss the algebraic structure we introduce, 
commutative monoids with cancellation. What is this structure, and why
do we introduce it? The answer is that this is precisely the structure
which describes the ``stringy'' aspect of quantum conformal field theory,
which means, on the closed worldsheet level, disjoint union and gluing of outbound
and inbound boundary components, and on the state space level usually some
type of tensor product and trace. This algebraic structure is perhaps somewhat
unusual to algebraists, in that it has ``dynamically indexed'' operations: instead
of a fixed set of operations, we have variable ``sets of inbound and outbound
boundary components'', and the gluing operations depend on these sets.
Algebraically, the best approach to this is to keep the concepts involved
as abstract as possible. This means, a commutative monoid with cancellation
consists of a commutative monoid $T$ (often, on the lax level which we are
interested in, it will simply be the category of finite sets), and for each pair
of elements $s,t\in T$ (``sets of inbound and outbound boundary components''),
a set $X_{s,t}$. The disjoint union operation is then a product
\beg{ecft2}{X_{s,t}\times X_{u,v}\r X_{s+u,t+v},
}
and the gluing is a ''unary'' operation of the form
\beg{ecft3}{X_{s+u,t+u}\r X_{s,t}.
}
A unit (``empty worldsheet'') $0\in X_{0,0}$ is also an operation (``constant'').
These operations satisfy obvious axioms (commutativity, associativity, distributivity),
which are described in detail in \cite{hk,hk1}. There, we also list the procedure
for obtaining coherence diagrams on the lax level. 

\vspace{3mm}
An important observation to make is that lax algebraic structures form a 
{\em $2$-category}. This means that we have the objects, $1$-morphisms,
which are functors laxly preserving the algebraic structure, and $2$-morphisms,
which are natural transformations of such functors, still compatible with the
algebraic structure. The simplest example of a $2$-category is the category
of categories (which is the same thing as ``lax sets''): objects then are categories,
1-morphisms are functors and 2-morphisms are natural transformations.

\vspace{3mm}
Let us now return to the stacks. Stacks are ``lax sheaves''. This means that we
must specify a site, which is a category $\mathcal{G}$, with a Grothendieck topology,
which means certain tuples of morphisms called coverings (generalizing colimits).
The category specifies objects $b$ over which we wish to index ``continuous families''.
The stack then specifies the ``category of sections'' over each object $b$.
A typical example for chiral CFT is the category of complex manifolds where coverings
are open covers. Sections over a complex manifold $b$ are then holomorphic
families of worldsheets, parametrized over $b$. In non-chiral cases (such as
physical CFT, or closed/open CFT), we do not have a notion of holomorphy, so
we must ``weaken'' the category $\mathcal{G}$. For the purposes of this paper,
we shall then consider just the category of real-analytic manifolds with
Grothendieck topology by open covers. 

\vspace{3mm}
Now given a site, stacks can be considered with values in any $2$-category which
has lax limits (see \cite{fi} for details). But this is true for any category
of lax algebras, such as lax commutative monoids with cancellation. This is
why the notion of SLCMC is possible. The main axiom of a stack is that it
be a lax contravariant functor form $\mathcal{G}$ into the $2$-category, which
takes Grothendieck covers to lax limits (a coherence diagram condition is also needed).

\section{The formalism for static brane theory}

\label{s2}

Again, this section is a review of the axiomatization of $D$-brane
categories given in 
\cite{hk}, and the reader should refer to that paper for a full discussion
of points which are only outlined here.

\vspace{3mm}
There are two boundary CFT formalisms discussed in \cite{hk}. In
the more general setup,
we discussed a notion of $D$-brane category where both the closed and
open sector can have a finite-dimensional anomalies, which means that
there is a finite-dimensional vector space of vacua. This means that
we had to introduce {\em labels} in both the open and closed string sector.
Labels in the open string sector are not the same as $D$-branes, rather,
each $D$-brane can have multiple labels. This ultimately led to a $3$-vector
space formalism for $D$-brane categories in \cite{hk}. 

\vspace{3mm}
From the
point of view of physics, however, the smaller the anomaly, the more interesting
the model is. In fact, ultimately, we would like the anomaly to vanish. This,
however, can be only discussed in the framework of full-fledged superstring theory.
In conformal field theory alone, the best we can hope for is {\em $1$-dimensional
anomaly} in both the open and closed sector. All the examples discussed in
the present paper will have $1$-dimensional anomaly in both sectors. 

\vspace{3mm}
Now $1$-dimensional anomaly allows us to decrease the level of category
theory involved by $1$, if we apply a ``fermionic''
approach to the anomaly, i.e. we consider the vacuum
as an element of the corresponding projective space
of the state space involvec. In the closed sector, 
then, no $2$-category theory is needed
(no labels); in the open sector, we can make do with a $2$-vector space
$\mathcal{B}$ of $D$-branes. To be more precise, let us first assume $\mathcal{B}$
is a finite-dimensional free $2$-vector space. 
Let us recall that a $2$-vector space is a lax module (in the sense outlined
in the previous section, see \cite{hk,fi} for details) over the lax commutative semiring
(which is the same thing as a symmetric bimonoidal category) of finite-dimensional
vector spaces and homomorphisms. One can then, by virtue of a general formalism
in $2$-category theory \cite{fi}, form lax functors such as $?\otimes_{\C_2}?$,
$Hom_{\C_2}(?,?)$. Since in CFT one needs to talk about Hilbert spaces,
one can also form the lax commutative algebra $\C_{2}^{Hilb}$ over $\C_2$
of Hilbert spaces, where
the product is the Hilbert tensor product. The notation $(?)^{Hilb}$ then
means $?\otimes_{\C_2}\C_{2}^{Hilb}$.

\vspace{3mm}
A finite-dimensional free $2$-vector space (which is what we assumed about 
$\mathcal{B}$ here) can be visualized simply as a product of finitely many
copies of $\C_2$: this is the ``alias'', or fixed base, interpretation of
the ``alibi'', or functorial definition we are giving here (see more comments
below).

\vspace{3mm}
Now we shall have a {\em closed
string state space} $\mathcal{H}$ and an {\em open
string state space} 
$\mathcal{K}\in Obj (\mathcal{B}\otimes_{\C_2}\mathcal{B}^{*})_{Hilb}$.
Here 
$$\mathcal{B}^*=Hom_{\C_2}(\mathcal{B},\C_2),$$
and the fact that $\mathcal{B}$ is finite and free implies a functor
$$tr:\mathcal{B}\otimes_{\C_2}\mathcal{B}^*\r\C_2.$$
(Note: it is a matter of discussion if the open sectors should be Hilbert spaces,
or if some other type of Banach space models them better;
however, we shall not go into that here.)

\vspace{3mm}
Now we define a closed/open CFT as a morphism of SLCMC's
\beg{edeff}{\mathcal{S}\r\mathcal{C}(\mathcal{B},\mathcal{K},\mathcal{H}).
}
Here all SLCMC's are stacks over the site of real analytic manifolds and
open covers. Over a point, the LCMC's have the underlying
lax commutative monoid the category of finite sets labelled by two
labels $closed, open$, i.e. maps of finite sets
$$S\r \{closed,open\}.$$ 
When referring
to an object $X_{b,c}$ of such SLCMC, we shall call $b$ resp. $c$ the set of inbound
resp. outbound boundary components. To make the definition \rref{edeff}
correct, there is one other subtlety (aside from the fact, that we, of course,
are yet to define the right hand side). The point is, there are different
types of arrangements of open and closed boundary components, and the
morphism \rref{edeff} must remember this data. This is taken care
of by defining \cite{hk} an auxiliary SLCMC $\Gamma$. For lack of a
better term, we call its objects {\em graphs}, although they are in fact
a rather special kind of graphs: these graphs are only allowed to have discrete
vertices and components which are circles. The discrete vertices
are labelled closed string inbound, outbound and brane, circles can have any number
of vertices $\geq 1$, are oriented, and their edges (corresponding to open string
boundary components) are additionally labelled inbound and outbound. 
(Note: what we refer to as $D$-brane boundary component is sometimes in
the literature referred to as free boundary.) Now it is clear how $\Gamma$
is an SLCMC over the same lax commutative monoid of sets with elements
labelled $inbound, outbound$, i.e. how the information it encodes behaves
under gluing (just imagine some abstract worldsheet with boundary described
by the graph). Consequently, we get an SLCMC \cite{hk}. Now the additional
requirement on \rref{edeff} is that both sides be equipped with morphisms
of SLCMC's into $\Gamma$, and that the morphism \rref{edeff} be {\em over $\Gamma$},
i.e. that the diagram formed by \rref{edeff} and the two morphisms into $\Gamma$
strictly commute.

\vspace{3mm}
Now the anomaly-free SLCMC $\mathcal{C}(\mathcal{B},\mathcal{K},\mathcal{H})$
over the SLCMC of graphs $\Gamma$ has sections over a point and over
$p_{in},p_{out}$ closed inbound and outboud components, $q$ pure $D$-brane
components, $s_{1},...,s_{\ell}$ open string components on $\ell$ 
mixed boundary components are points of
\beg{eoc1}{\cform{\bigotimes}{i=1}{p_{in}}\mathcal{H}^*
\otimes \cform{\bigotimes}{i=1}{p_{out}}\mathcal{H}
\otimes
\cform{\bigotimes}{i=1}{\ell}
tr_{cyc}\left(\cform{\bigotimes}{j=1}{s_{i}}\mathcal{K}^{(*)}\right)}
Here we supress completion from the notation (always taking
trace class elements in the sense of \cite{hk} in the Hilbert
tensor product), and the ${}^{(*)}$ superscript
means that dual is assigned to those string components which correspond to
inbound open string components. 
The symbol $tr_{cyc}$ means
that we take trace as many times as we have $D$-brane components on each
mixed boundary component; in the end, \rref{eoc1} is just a Banach space
(subset of a Hilbert space), and doesn't have any labels.
Actually, the dual in the open sector
requires further discussion. Let us, for simplicity, work with objects
of $2$-vector spaces, the Hilbertization case follows analogously.
Assume, therefore, we have
\beg{eoc2}{x\in Obj(\mathcal{B}).}
We claim that we have a canonical object
\beg{eoc3}{x^*\in Obj(\mathcal{B}^*)}
together with a canonical morphism
\beg{eoc4}{x\otimes x^*\r 1}
where 
$$1\in Obj\mathcal{B}\otimes \mathcal{B}^*\simeq Hom(\mathcal{B},\mathcal{B})$$
is the identity (note that the second equivalence follows from the assumption
that $\mathcal{B}$ is finite and free).

\vspace{3mm}
Assuming for the moment that for \rref{eoc2}, we have \rref{eoc3} with
\rref{eoc4}, by $\mathcal{K}^*\in Obj(\mathcal{B}\otimes
\mathcal{B}^*)$ we mean the coordinate-wise dual; this would lie in
$Obj(\mathcal{B}^*\otimes\mathcal{B})$, so we switch the coordinates,
so they appear in the same order as in the inbound components:
$Obj(\mathcal{B}\otimes\mathcal{B}^{*})$. Then, we can take the cyclic
trace regardless on which open string components are inbound and which
are outbound.

\vspace{3mm}
The sections \rref{eoc1} are ``stacked'' in the usual way, we skip that discussion
(see \cite{hk}).
The only thing to point out is that, as usual, open worldsheets do not form
a complex manifold, so we cannot have stacks of holomorphic sections, the
best we can do is {\em real-analytic} sections.

\vspace{3mm}
A note is however in order on gluing of the sections \rref{eoc1}.
That requires the following additional construction. Consider the following
diagram:
\beg{eoc5}{{
\diagram
\mathcal{B}\drdashed & & & \mathcal{B}^{*}\dldashed\\
&\mathcal{B}^*\dline^{\mathcal{K}}& \mathcal{B}\dline_{\mathcal{K}^{*}} &\\
& \mathcal{B}\dldashed & \mathcal{B}^*\drdashed &\\
\mathcal{B}^{*}&&&\mathcal{B}.
\enddiagram
}}
The diagram is to be read as follows: Along each solid line, we take the
object denoted, and along each dotted line, we take the trace (=evaluation).
We need to explain how the diagram \rref{eoc5} maps naturally to the diagram
\beg{eoc6}{{
\diagram
\mathcal{B}\rrdashed &&\mathcal{B}^*\\
\\
\mathcal{B}^{*}\rrdashed &&\mathcal{B}.
\enddiagram
}}
Note that by the triangle identity, 
$$\diagram
\mathcal{B}\rdashed&\mathcal{B}^*\rline^{1}&\mathcal{B}\rdashed&\mathcal{B}^*
\enddiagram
$$
is canonically isomorphic to 
$$\diagram
\mathcal{B}\rdashed &\mathcal{B}^*.
\enddiagram$$
Therefore, it suffices to exhibit a canonical map from
$$\diagram
\mathcal{B}^*\dline^{\mathcal{K}}&\mathcal{B}\dline_{\mathcal{K}^*}\\
\mathcal{B}&\mathcal{B}^*,
\enddiagram$$
into
$$\diagram
\mathcal{B}^*\rline^{1}&\mathcal{B}\\
\mathcal{B}\rline^{1}&\mathcal{B}^*
\enddiagram
$$
but that follows from \rref{eoc4}.

\vspace{3mm}
To construct, from \rref{eoc2}, \rref{eoc3} with \rref{eoc4}, we consider
$x$ as a functor
$$x:\C_2\r \mathcal{B},$$
and let
$$x^*:\mathcal{B}\r \C_2$$
be its right adjoint. Recall here that a functor $G$ is right adjoint to $F$
or equivalently $F$ is left adjoint to $G$ if we have a natural
bijection
\beg{eadj}{Hom(x,Gy)\cong Hom(Fx,y).} 
\rref{eoc4} then follows from the so called triangle identities for
adjoint functor, which are direct consequences of \rref{eadj}.
To be more specific, the counit of the adjunction is of the form
$$x\otimes x^*(y)\r y$$
which, under the equivalence 
\beg{eoc10}{\mathcal{B}\otimes\mathcal{B}^*
\simeq Hom_{\C_2}(\mathcal{B},
\mathcal{B})} 
is
$$x\otimes x^*\r Id,$$
which is \rref{eoc4}.

\vspace{3mm}
As pointed out above, in order to have CFT examples, we need to allow $1$-dimensional
anomaly. This is done by introducing the SLCMC
$$\tilde{\mathcal{C}}(\mathcal{B},\mathcal{K},\mathcal{H})$$
whose sections over a point consist of a complex line $L$ and a linear
map from $L$ to the space of sections of 
$\mathcal{C}(\mathcal{B},\mathcal{K},\mathcal{H})$. 
Since gluing is linear,
this behaves well with respect to gluing. Sections over an open set in the
category of real-analytic manifolds are again defined in the usual way
(see \cite{hk}).
(There is also an
adjoint approach where one sticks to the non-anomalous SLCMC in the target,
and uses a ``$\C^{\times}$-central extension'' of the source SLCMC, cf. 
\cite{hk1}.)

\vspace{3mm}

Up to now in this section, and also in
\cite{hk}, we have tried to express the formalism for
closed and open CFT's in ``alibi'' form, with as much functoriality as possible.
However, for various reasons, it is also useful to have an ``alias'' interpretation,
i.e. to express everything in terms of bases. In the current setting,
this means that we write $\mathcal{B}$ explicitly as a free $2$-vector space
on a finite set of ``elementary $D$-branes'' $B$. We can then think of
the open state space as simply a set of Hilbert spaces
\beg{eal1}{\mathcal{K}_{ab}}
of states of outbound open string beginning on the $D$-brane $a$ and ending
on the $D$-brane $b$. One finds that this is not necessarily symmetric
in $a$ and $b$. The state space of an inbound open string whose string beginning
point is on $a$ and string endpoint is on $b$ is then
\beg{eal2}{\mathcal{K}_{ab}^{*}.
} 
This is easily seen to coincide with the above interpretation; in particular,
it makes gluing work. 

\vspace{3mm}
One may wonder why introduce the more complicated ``alibi'' interpretation
above. The answer is that it is appealing to follow the philosophy of
``replacing sets with $k$-vector spaces with $k\in\{0,1,2,...\}$ as low as 
possible. In mathematics, it is known that a stable theory $k$-vector spaces
is ``$v_k$-periodic'' \cite{bdr}. For $k=0$, it means that the theory is
a kind of ordinary cohomology, for $k=1$ a kind of $K$-theory, for $k=2$
a kind of elliptic cohomology. The functoriality we detect therefore 
predicts that ``charges'' of $D$-branes in a $D$-brane category lie
in a kind of $K$-theory, as discovered by Witten (cf. \cite{w}). The functoriality
discovered in \cite{hk} for example
predicts that a $D$-brane category with higher-dimensional
anomaly (a modular functor associated with open string sectors) has charges
in a kind of elliptic cohomology.

\vspace{3mm}
Unfortunately, this kind of connection is hard to make precise without
somehow ``group-completing'' the category of $k$-vector spaces. This is because
for $k\geq 2$, the category itself has too few isomorphisms, as is well known
(cf. \cite{bdr}). In \cite{hk}, we propose a way of group-completing the
category of vector spaces using topology. This predicts that modular functors
with values in super-vector spaces will be needed to fully understand the
connection of CFT's with elliptic cohomology. One already knows that in the
open sector, such formalism will be needed to fully understand anti-$D$-branes.

\vspace{3mm}
In this paper, we do not discuss the group completion of the category of vector
spaces, but topology does come into play when we introduce a mathematical
formalism for the Kondo effect in WZW branes.

\vspace{5mm}

\section{An example: Cardy branes}

\label{s2a}

Cardy's theory \cite{cardy} considers an RCFT, which is a conformal field
theory of the form 
\beg{ecard1}{\mathcal{H}=\cform{\bigoplus}{\lambda}{}\mathcal{H}_{\lambda}
\otimes\overline{\mathcal{H}}_{\lambda}.
} 
Here the tensor product is Hilbert tensor product, and $\overline{\mathcal{H}}$ 
denotes the complex conjugate
Hilbert space (which is canonically isomorphic
to the dual Hilbert space $\mathcal{H}^*$). 
The collection of HIlbert spaces $\mathcal{H}_{\lambda}$ is a closed chiral
CFT in alias notation, i.e. the $2$-vector space of labels is
$$\mathcal{M}=\cform{\bigoplus}{\lambda}{}\C_2.$$
Recall from \cite{hk} that our formalism for modular functor
and closed chiral CFT is given by morphisms of SLCMC's
$$\mathcal{C}\r \mathcal{C}(\mathcal{M}),$$
$$\mathcal{C}\r \mathcal{C}(\mathcal{M},\mathcal{H}),$$
respectively, where $\mathcal{C}$ is the SLCMC of closed worldsheets
(over the site of complex manifolds and open covers; over a point,
in the closed case, all SLCMC's can be taken to have as the underlying
lax commutative monoid the category of finite sets. It becomes
more precisely covering spaces in the stack context). Here $\mathcal{M}$
is a finite free $2$-vector space of {\em labels}. The category of sections of
the SLCMC $\mathcal{C}(\mathcal{M})$ over a point and over sets of $p$ inbound and
$q$ outbound components is
$$\mathcal{M}^{*\otimes p}\otimes\mathcal{M}^{\otimes q}.$$
Now choose $\mathcal{H}\in Obj(\mathcal{M}^{Hilb})$.
To define the SLCMC $\mathcal{C}(\mathcal{M},\mathcal{H})$, we let the
sections over a point and over $p$ inbound and $q$ outbound elements
be the category of sections $M$ of $\mathcal{C}(\mathcal{M})^{p,q}$ together
with trace-class morphisms
$$M\r  \mathcal{H}^{*\otimes p}\otimes\mathcal{H}^{\otimes q}.$$
Here the tensor product on the right means Hilbert tensor product, and
trace class means that there exist bases in all tensor factors with
respect to which, when we expand the given element, we obtain a convergent
norm sum (as opposed to just quadratic-convergent). See \cite{hk} for
details.

\vspace{3mm}
We may then consider 
$$(\mathcal{H}_{\lambda})\in Obj(\mathcal{M}^{Hilb})$$
as in \cite{hk}. We assume that the 
the modular functor satisfies, in the alias notation, 
\beg{ecard2}{M_{\overline{X}}\cong \overline{M_X}}
(where $M_X$ is the value of the modular functor on a worldsheet $X$)
subject to the condition that \rref{ecard2} is an
isomorphism of modular functors; here $X$ is a closed worldsheet
with labelled boundary components, and $\overline{X}$ is the complex-conjugate worldsheet
with labels replaced by contragredient labels. We also assume that there
is a condition on
vacua. 
Next, we also require ``reflection
positivity'', i.e. that, in the alias notation, for $\mu\in M_X$,
\beg{ecard4}{U_{X,\mu}=\overline{U_{\overline{X},\mu}}
}
where $U_{X,\mu}$ is the vacuum of the labelled worldsheet $X$ assigned to
a given element of the modular functor $M_X$. (Note that in $\overline{X}$,
the outbound boundary components of $X$ become inbound and vice
versa, which makes the target Hilbert spaces on both sides of
\rref{ecard4} dual, hence complex conjugate.) In \rref{ecard1},
we consider $M_{\overline{X}}$, $\overline{\mathcal{H}}_{\lambda}$ functors not in $
\overline{X}$
but in $X$, thereby making them antiholomorphically dependent
on $X$ (or `antichiral', or `right-moving'). To make \rref{ecard1} a CFT
with $1$-dimensional anomaly, more is however needed. Namely, one needs
a non-degenerate pairing
\beg{ecard5}{M_X\otimes \overline{M_X}\r L_X
}
where $L_X$ is a $1$-dimensional modular functor,
subject to the condition that
\rref{ecard5} is a morphism of modular
functors (cf. \cite{scft}). 

\vspace{3mm}
It is conjectured that the level $k$ WZW model
on $SU(n)$ is an example. Although a proof
has not been published, the authors 
believe that a proof is essentially contained
in the literature. Looking at the chiral part, for example
a vertex operator algebra version of this theory
has been constructed by Frenkel and Zhu \cite{fzhu}.
The modular functor, in their own formalism, has been
treated in detail by Bakalov-Kirillov \cite{bk}.
The main question is whether the worldsheet vacua converge 
in Segal's formalism which we use here. However, the 
level $k$ WZW model is a subquotient of a 
free field theory by \cite{gh}. For free field theories,
it is known that the scaled vertex operators corresponding
to worldsheets are smooth (cf. \cite{hk1}),
i.e. decay exponentially with weight. Using gluing, this can be used to
show that the worldsheet vacua in fact converge in the sense of the
Segal model. Nevertheless, all the details related mostly
to compatibility of the various approaches to CFT, have not been checked,
and we do not prove this conjecture here. 

\vspace{3mm}
In the case of the level $k$ WZW model for $SU(n)$, 
the labels are indexed by $\lambda\in P_k$,
where $P_k$ is a set enumerating level $k$ irreducible lowest weight
representations of the
universal central extension $\tilde{L}SU(n)$ of the loop group $LSU(n)$,
and $\mathcal{H}_{\lambda}$ is the representation corresponding to $\lambda$.
We refer the reader to \cite{ps} for details on loop groups.

\vspace{3mm}

Now Cardy \cite{cardy,sch,fs,mms,moore} argues that an RCFT always
has the following $D$-brane category:
the $2$-vector spaces $\mathcal{B}$ of
branes is equal to the $2$-vector space
of the closed labels $\lambda$ of its chiral theory. $\mathcal{B}$ is
free on a set of ``elementary branes'' $B$ (in the case of the WZW model,
$B=P_k$). Furthermore, using the formalism of boundary states, Cardy states
that the open sectors in alias notation should be
\beg{ecb1}{\mathcal{K}_{\lambda\mu}=\cform{\bigoplus}{\kappa}{}N_{\lambda\mu^*}^{\kappa}
\mathcal{H}_{\kappa}.
}
Here $\mu^*$ denotes the contragredient label to $\mu$, $N_{\alpha\beta}^{\gamma}$
is the fusion rule, i.e. the dimension of the chiral modular functor on a pair
of pants with two inbound boundary components labelled by $\alpha, \beta$ and
one outbound boundary component labelled by $\gamma$.

\vspace{3mm}
Again, however, a construction of the open vacua in Segal's approach has
not been published in the literature, although related constructions are
given in \cite{bcdcd,fffs}: in \cite{bcdcd}, the construction of vacua
is approached physically in the language of Polyakov path integrals
and string determinants, \cite{fffs} relates rigorously mathematically
boundary charges to $3$-dimensional topological quantum field theory
(following the closed worldsheet treatment of \cite{tur}). \cite{fffs},
however, does not discuss CFT vacua.
The construction of the open vacua can be outlined as follows: Consider
an open worldsheet $\Sigma$. First, assume for
simplicity that there are no closed string boundary components. 
Although the $D$-brane components are not parametrized, we can nevertheless use
them to glue canonically $\Sigma$ to its complex conjugate $\overline{\Sigma}$,
thus turning it into a closed worldsheet $\Sigma\overline{\Sigma}$ (the $D$-brane
boundary components are used up in the gluing, the open string components
turn into closed ones). 
To be more precise, if we denote by $\partial_{b}(\Sigma)$ the union of the $D$-brane
components of $\Sigma$, then there is a canonical diffeomorphism 
\beg{ecand}{\partial_{b}(\Sigma)\cong \partial_{b}(\overline{\Sigma})
}
(recalling the construction of the complex conjugate complex manifold, 
the underlying manifold of $\overline{\Sigma}$ can in fact be taken to be equal
to the underlying manifold of $\Sigma$, in which case the map \rref{ecand} is
the identity). Thus, \rref{ecand} provides a canonical recipe for gluing
$\Sigma$ and $\overline{\Sigma}$. Now specifying the gluing map endows
$\Sigma\overline{\Sigma}$ with a canonical complex structure. In the
case of analytic boundary components (which we can assume here), this
is an elementary direct construction, which can be extended to non-analytic
smooth boundary components (and even further) by Ahlfors-Bers theory:
this is the
basic gluing operation of Segal's approach to CFT, although we are using
it in a different way here, gluing the brane rather than string components
on the boundary.

\vspace{3mm}
{\bf Remark:} A reader familiar with \cite{fffs} may find this confusing,
since the construction given there appears slightly different. However,
the point is that \cite{fffs} proceeds in the language of correlation
functions, which means that string components of the boundary are not
present at all, and are replaced by punctures, which are labelled points on the
worldsheet. Thus, the entire boundary of the worldsheet as considered
in \cite{fffs} consists in fact of brane components. A labelled point
in the interior of the worldsheet in \cite{fffs} corresponds in our language to a closed
string component, a labelled point in on the boundary in \cite{fffs}
corresponds to an
open string component. With this translation, our construction is exactly
the same as the doubling construction of \cite{fffs} in the limit where
string components degenerate to points. Even more directly, one
can relate our construction to that of \cite{fffs} without taking limits
by gluing (in our sense, following
Segal) standard disks (resp half-disks) with labelled point in the origin
to all closed (resp. open) string components of our worldsheet, thus
obtaining a worldsheet in the sense of \cite{fffs}. This operation then
converts our doubling construction exactly to the doubling construction
of \cite{fffs}.

\vspace{3mm}
In any case, this gives us a map
\beg{ecb2}{M_{\Sigma\overline{\Sigma}}\r \bigotimes \mathcal{H}_{\kappa_i}^{(*)}.
}
Here $\kappa_i$, $i=1,...,s$, are labels ($\in\mathcal{B}$) 
of the string components of $\Sigma$
(also $\Sigma\overline{\Sigma}$), ${}^{(*)}$ as usual denotes 
dualizing wherever an inbound string component occurs, and $M$, as above, is
the modular functor of the corresponding chiral theory.

\vspace{3mm}
The idea is to use the map \rref{ecb2} for defining open string vacua, and to
show that when we take the particular linear combinations \rref{ecb1} of
the $\mathcal{H}_{\kappa}$'s, the source of \rref{ecb2} can somehow be reduced
canonically, compatibly with gluing, to a $1$-dimensional subspace.
To this end, it is quite clear that we shouldn't interpret the fusion coefficients
in \rref{ecb1} as numbers, but as actual values of the chiral modular functor
corresponding to a particular (``reference'')
pair of pants. Actually, we don't want an arbitrary
pair of pants, we want it to be of the form
$P\overline{P}$, i.e. sewn together from
two complex conjugate ``front'' and ``back'' pieces, each of which is an open
worldsheet $P$ of genus $0$ with one boundary component and three string components,
labelled by $\kappa$, $\lambda$, $\mu$. The components labelled by
$\kappa$, $\mu$ have the same orientation, the component labelled by
$\lambda$ opposite. In our case, we will therefore have
$s$ such parts $P_i$, $i=1,...,s$. Here $P_i$ has one string component $c$
labelled by $\kappa_i$ of orientation opposite to $\kappa_i$
(so they can be glued), and
two string components $d,e$ 
which are labelled by the $D$-branes of $\Sigma$ to which they
are adjacent. Let us assume the $d$ component is adjacent to the brane
of $\Sigma$ which comes before $c$ in the clockwise order. Then
$d$ has the orientation opposite to $c$ in $P_i$, and $e$
has the same orientation as $c$. 

\vspace{3mm}
Now sew the open worldsheets $\Sigma$, $P_i$, $i=1,...,s$ to obtain a new
worldsheet $\Theta$, and then sew the front and back to obtain a closed
worldsheet $\Theta\overline{\Theta}$. This worldsheet $\Theta\overline{\Theta}$
now has closed string components, labelled by the $D$-branes $\lambda$ of $\Sigma$. 
In fact, each $D$-brane component $\alpha$ of $\Sigma$ corresponds to {\em two}
boundary components $c_{\alpha}$, $\overline{c}_{\alpha}$
of $\Theta\overline{\Theta}$ with the same label $\lambda_{\alpha}$ as
the $D$-brane corresponding to $\alpha$, of opposite orientations. Also,
it is symmetrical with respect to complex conjugation. We want to show that
the modular functor $M_{\Theta\overline{\Theta}}$ contains a canonical line. 

\vspace{3mm}
The idea is that although $c_{\alpha}$, $\overline{c}_{\alpha}$ are on the
``plane of symmetry'' of $\Theta\overline{\Theta}$ (the symmetry
being complex conjugation), since they have opposite
orientations, they can be moved to opposite sides (into the interior of 
$\Theta$, $\overline{\Theta}$ without breaking the symmetry of $\Theta\overline{\Theta}$.
In other words, the worldsheet $\Theta\overline{\Theta}$ is perturbed
into a new worldsheet $\Theta^{\prime}\overline{\Theta^{\prime}}$ where
now $\Theta^{\prime}$ is a closed worldsheet. (Further, the perturbation move
can be expressed by tensoring with a known line, since we know the modular
functor of cylinders.)

\vspace{3mm}
Now in alias notation, let $\gamma$ be a set of labels to put on the new closed
string components of $\Theta^{\prime}$, via which it is
attached to $\overline{\Theta^{\prime}}$ (the other string components are already
labelled). Then we have
\beg{ecb10}{M_{\Theta^{\prime}\overline{\Theta^{\prime}}}=\cform{\bigoplus}{\gamma}{}
M_{\Theta^{\prime}_{\kappa}}\otimes \overline{M_{\Theta^{\prime}_{\kappa}}}}
by \rref{ecard2}. But the right hand side of \rref{ecb10} contains a canonical
line by \rref{ecard5}.

\vspace{3mm}
The case when the original worldsheet $\Sigma$ has closed string components is
handled in a trivially modified manner: one repeats verbatim the construction of 
the worldsheet $\Theta$, and uses the vacuum in the chiral theory. If there
are $p$ closed string components, then
the state space of the worldsheet $\Sigma\overline{\Sigma}$
picks up a factor 
$$\cform{\bigotimes}{i=1}{p}\mathcal{H},$$
(summing over all the possible choices of labels on the closed string
components, which we require to be the same in $\Sigma$ and $\overline{\Sigma}$).
Additionally, the modular functor of $\Sigma\overline{\Sigma}$
(and also $\Theta\overline{\Theta}$)
picks up an additional factor of
\beg{ecb11}{\cform{\bigotimes}{i=1}{p} \left(\cform{\bigoplus}{\lambda\in B}{}
M_{\lambda}\otimes \overline{M_{\lambda}}
\right)
}
which, as we know from \rref{ecard5}, again contains a canonical line. Compatibility
under gluing is not difficult to verify.

\vspace{3mm}

Another subtlety in the above discussion is the case of closed $D$-brane
components $c$. Strictly speaking, the construction we just described only gives a
{\em sum} of states over all $D$-brane labels in this case, because the labelling
of the closed $D$-brane component disappears during the doubling. To remedy this
in a way which is most convenient for proving consistency under gluing, we may
cut $c$ by  
a short open string component $s$, and endow it with a reference half-pair of
pants as before. The labels this way will be retained in the doubling, and we
may recover the state corresponding to $c$ by gluing back in the open string vacuum
corresponding to $s$. This is consistent, since we know how to move strings
around in $\Theta\overline{\Theta}$ using the Virasoro algebra, and $s$ can
always be moved out of the way of any cuts to prove consistency under gluing.

\vspace{3mm}

However, there is a better approach which may be
preferrable calculationaly: we may replace the
closed $D$-brane component by a closed string component, and then glue in the
``boundary state'', corresponding to the cylinder $A_{\tau}$ with one closed $D$-brane
and one closed string component (parametrized in the standard
way), obtained by gluing a parallelogram $Q$
with one side $1$ and one side $\tau$ in $\C$. 
It is convenient to assume that $\tau=0$, i.e. the cylinder
$A_{0}$ has width $0$, i.e. is degenerate. This will cause the state to be divergent
from in the Hilbert space (i.e. distributional), but that doesn't matter: we may 
always glue in the vacuum of a string annulus to get a true trace
class Hilbert state.
Then the boundary state corresponding to brane $\lambda\in B$ must have the form
\beg{eis1}{b_{\lambda}=\cform{\sum}{\mu\in B}{}\alpha_{\lambda\mu} 1_{\mu}}
where $1_{\mu}:\mathcal{H}_{\mu}\r \mathcal{H}_{\mu}$ is the identity
(called the `Ishibashi state', see \rref{ecard1},
\cite{mms}, Section 5.1), and $\alpha_{\lambda\mu}$
are some coefficients. 

\vspace{3mm}
The trick for calculating the coefficients
$\alpha_{\lambda\mu}$ (see e.g. \cite{mms}, \cite{sch})
is to actually consider the cylinder $A_{\tau/2}$
with finite width $\tau/2$ and glue it with the opposite cylinder $A_{\tau/2}^{\prime}$
where the string boundary component is oriented the opposite way. Let $C_{\tau}$
be the cylinder with two $D$-brane components obtained by gluing $A_{\tau/2}$
and $A_{\tau/2}^{\prime}$, with $D$-brane component labelled
by another $D$-brane $\lambda^{\prime}$. Then, using the usual anomaly trivialization on
rigid cylinders, the vacuum corresponding to $C_{\tau}$ is
\beg{eis2}{U_{A_{\tau}}=\cform{\sum}{\mu\in B}{}\alpha_{\lambda\mu}
\alpha_{\lambda^{\prime*}\mu}Z_{\mu}(q)
}
where $Z_{\mu}$ is the partition function of the chiral sector $\mh_{\mu}$,
and $q=e^{2\pi i \tau}$. 

\vspace{3mm}
On the other hand, however, $U_{A_{\tau}}$ can be computed by cutting
$A_{\tau}$ into a parallelogram $Q$ with two open $D$-brane components 
parallel to the vector $\tau\in\Z$, and two open string components parallel
to the vector $1\in \C$ which is conformally isomorphic to 
the parallelogram where the open $D$-brane components are parallel to
$1$ and the open string components are parallel to $\tau^{\prime}=-1/\tau$.
We can calculate this trace using the ``doubling'' described above.
The expression one obtains is
\beg{eis3}{tr U_{q}=\cform{\sum}{\mu\in B}{}N_{\lambda\lambda^{\prime *}}^{\mu} 
Z_{\mu}(q^{\prime})
}
where $q^{\prime}=e^{2\pi i \tau^{\prime}}$.
Now consider the modular $S$-matrix, i.e. the unitary matrix corresponding
to the modular transformation $-1/\tau$, in the basis $B$ (the elementary
closed labels). Then, using the Verlinde conjecture \cite{v} (the proof of
which in the present context is outlined in \cite{scft}, an earlier proof
in more
physical setting appearing in \cite{ms}), we compute
\beg{eis4}{{
\begin{array}{l}
\cform{\sum}{\mu}{}  N_{\lambda\lambda^{\prime*}}^{\mu}Z_{\mu}(q^{\prime})=\\
\cform{\sum}{\kappa}{}\fracd{S^{\kappa}_{\lambda}S^{\kappa}_{\lambda^{\prime*}}
(S^{-1})^{\kappa}_{\mu}}{S^{\kappa}_{0}}Z_{\mu}(q^{\prime})=\\
\cform{\sum}{\kappa}{}\fracd{S^{\kappa}_{\lambda}S^{\kappa}_{\lambda^{\prime*}}}{
S^{\kappa}_{0}}Z_{\kappa}(q)
\end{array}
}}
(where $0$ is the zero label). Now plugging \rref{eis4} into \rref{eis3} and
using the equality with \rref{eis2}, we obtain the equations
\beg{eis5}{\alpha_{\lambda\mu}
\alpha_{\lambda^{\prime*}}=\fracd{S^{\mu}_{\lambda}S^{\mu}_{\lambda^{\prime*}}}{
S^{\mu}_{0}},
}
which has, up to possible multiple of $-1$, a unique solution
\beg{eis6}{\alpha_{\lambda\mu}=\fracd{S^{\mu}_{\lambda}}{\sqrt{S^{\mu}_{0}}}}
(for example set $\lambda^{\prime*}=\lambda$).
Plugging back into \rref{eis1} gives the formula \cite{sch,mms} for the boundary
state,
\beg{eis7}{b_{\lambda}=\cform{\sum}{\mu\in B}{}\fracd{S^{\mu}_{\lambda}}{
\sqrt{S^{\mu}_{0}}} 
1_{\mu}.
}

\vspace{3mm}
A note is due on the status of the claim that the level $k$
$SU(n)$-WZW model is an RCFT: as far as we know, a rigorous proof 
written in the present setting is not
anywhere in the literature. However, it seems that
a rigorous proof can be obtained by combining the results of Bakalov
and Kirillov \cite{bk} with the technique for proving convergence
results by the boson-fermion correspondence \cite{hk1}. Therefore,
modulo writing down certain tedious details, the WZW RCFT, and 
its theory of Cardy branes, is on fairly solid ground.

\vspace{3mm}
Additionally, it is worth pointing out that in the WZW model
of a compact Lie group $G$, the brane
vacua have the following geometrical interpretation. For simplicity,
instead of the Cardy vacua, we consider the ``Ishibashi vacua'', i.e.
open worldsheets where branes are unlabelled and open strings are
labelled by $\lambda\in P_k$. It is then appropriate to consider the
state space $\mh_{\lambda}$ as a representation of a particular loop
group as follows: the closed state space $\mh$ is a representation
of a central extension of the loop group
\beg{el1}{LG\times LG.}
Consider now the unit disk $D$ and let its boundary be the source of the
loops in \rref{el1}. Consider now the open worldsheet $D^+=
\{z\in D|Im(z)\geq 0\}$ where the $D$-brane component is $D^+\cap\R$.
We may then consider the Ishibashi open string sector $\mh_{\lambda}$
as an irreducible representation of the subgroup 
\beg{el2}{LG_{\Delta}=\{(f,g)\in LG\times LG|
f(z)=g(\overline{z}) \;\text{for $z\in S^1$}
\}
}
of \rref{el1}. There are other variations of this definition, for example
we may look at the group $LG_{o}$ of pairs of functions $(f,g)$ from a neighborhood
$U$ of $S^1\cap D^+$ into $G_{\C}$ where $f$ is holomorphic, $g$ is antiholomorphic,
and $f\equiv g$ on the real line. The advantage of the definition of
$LG_{o}$ is that it is more ``local''. Indeed, for an Ishibashi open worldsheet
$\Sigma$, (which, for simplicity, we assume contains no closed string
components), we may look at the subgroup
\beg{el3}{
LG_{\Sigma}\subset\prod LG_{0}
}
(the product is over boundary components) given by pairs $(f,g)$ of functions
$$\Sigma\r G_{\C}$$
where $f$ is holomorphic, $g$ is antiholomorphic and 
$f\equiv g$ on $D$-brane components of $\Sigma$. 
Then using standard methods (actually in this case easier for the WZW model
than the lattice CFT), one shows that the cocycle defining the universal central
extension of $\prod LG_{o}$ vanishes when restricted to $LG_{\Sigma}$.
The Ishibashi vacuum on $\Sigma$ is then defined as the fixed subspace
\beg{el4}{(\prod \mh_{\kappa_{i}})^{LG_{\Sigma}}}
of the state space. Of course, \rref{el4} agrees with the above ``doubling''
construction for general RCFT.

\vspace{5mm}

\section{Brane dynamics: infinitesimal deformations}

\label{s3}

It is well known that a closed CFT with $1$-dimensional anomaly can be
infinitesimally deformed by integrating a fixed primary field of weight $(1,1)$
over the worldsheet
(see \cite{scft, hk1}). The Kondo effect, in its abstract mathematical
form, is a similar
construction allowing us to modify infinitesimally
the open sector of a fixed closed CFT 
$\mathcal{H}$. To be more precise, suppose we have an closed/open CFT
\beg{ek1}{\Phi:\mathcal{D}\r \tilde{\mathcal{C}}(\mathcal{B}, 
\mathcal{K},\mathcal{H})
}
where $\tilde{\mathcal{C}}(\mathcal{B}, 
\mathcal{K},\mathcal{H})$
is as in Section \ref{s2} as above, 
and $\mathcal{D}$ is the SLCMC of closed/open worldsheets as in \cite{hk}.

\vspace{3mm}
Now suppose we have a brane $a\in Obj(\mathcal{B})$. Consider
$a^*\in Obj(\mathcal{B}^*)$ as in Section \ref{s2} above. Then we may consider
\beg{ek2}{a\otimes a^*\in Obj(\mathcal{B}\otimes \mathcal{B}^*). 
}
But we have
\beg{ek3}{\mathcal{B}\otimes \mathcal{B}^*\simeq Hom_{\C_2}(
\mathcal{B}\otimes \mathcal{B}^*,\C_2),
}
so we may also consider $a\otimes a^*$ as a morphism of $2$-vector spaces:
\beg{ek4}{a\otimes a^*:\mathcal{B}\otimes \mathcal{B}^*\r\C_2.
}
This way, we may define the open sector
\beg{ek5}{\mathcal{K}_{aa}=(a\otimes a^*)(\mathcal{K}).
}
Now the semigroup $\C^{\times<1}$ of rigid annuli doesn't act on $\mathcal{K}_{aa}$,
but its subsemigroup $(0,1)=\R^{\times<1}_{+}$ (the ``renormalization group'') does.
An element of $\mathcal{K}_{aa}$ is said to have {\em weight $k$} if
$t\in (0,1)$ acts by $t^k$ on $x$.
In fact, more generally, we shall consider the semigroup $\mathcal{G}$
of holomorphic maps $f:D\r D$ (where $D$ is the unit disk in $\C$) with
the property that $f(0)=0$, $f([0,1])\subseteq [0,1]$, 
$f([-1,0])\subseteq [-1,0]$. Note that every such map $f$ necessarily has
real derivative at $0$. Note also that such map necessarily defines an open
worldsheet, and hence a map
\beg{ek6}{f:\mathcal{K}_{aa}\r\mathcal{K}_{aa}.}
(The central extension necessarily splits canonically on worldsheets of this
type, similarly as in the closed sector.) We shall call $x\in \mathcal{K}_{aa}$
a {\em primary field of weight $k$} if for every $f\in\mathcal{G}$,
\beg{ek7}{f(x)=f^{\prime}(0)^{k}\cdot x.
}
We claim that then it is possible to construct an {\em infinitesimal deformation}
$a_x$ of the brane $a$ along the element $x$. This is more easily explained
in the ``alias'' interpretation, i.e. with the choice of a closed/open worldsheet
$\Sigma$ with $p$ closed string, $q$ closed $D$-brane and $s$ mixed boundary
components. Assume the $m$'th mixed boundary component, $m=1,...,s$, has $\ell_m$
open $D$-brane components, to which we assign $D$-branes $a_{mi}$,
$i=1,...,\ell_m$ (we shall set $a_{m0}=a_{m,\ell_{m}}$). Assume also
to the $j$'th closed $D$-brane boundary component we assign the $D$-brane
$a_{j}$. Then for this
choice $c$, our formalism
gives us a vacuum vector
\beg{ek8}{U_{c}\in \cform{\bigotimes}{i=1}{p}\mathcal{H}^{(*)}\otimes
\cform{\bigotimes}{m=1}{s}\left(
\cform{\bigotimes}{i=0}{\ell_m-1}\mathcal{K}_{a_{mi}a_{m,i+1}}^{(*)}
\right)
}
where, as usual, ${}^{(*)}$ denotes dual if the corresponding string component
is inbound, and notation for Hilbert-completing the tensor products and
then passing to trace class elements is omitted from the notation.

\vspace{3mm}
The infinitesial deformation is described as follows: suppose that on finitely
many $D$-brane boundary components $c_1,...,c_r$ of $\Sigma$, (open or closed),
the $D$-brane assigned to each $c_i$ in the choice $c$ is $a$. (But
we allow $a$
to be also possibly assigned to other $D$-brane components of $\Sigma$.)
Then we may define a deformed choice $c'$ which is the same
as $c$ except that the $D$-brane
assigned to each $c_i$ is the ``infinitesimally deformed $D$-brane $a_x$''.
The ``vacuum deformation'' is defined simply by
\beg{edef1}{\Delta U_c=
\int_{c_1\cup...\cup c_r} \Psi_t(x).
}
This is to be read as follows: $t$ is an arbitrary point of $c_1\cup...\cup c_r$.
The element $\Psi_t(x)$ is defined by choosing a holomorphic embedding $g:D^{+}\r \Sigma$
where 
$$D^+=\{z\in\C|\;||z||\leq 1,Im(z)\geq0\}$$ 
such that $g(0)=t$. Consider a new worldsheet $\Sigma_{g}$ obtained from $\Sigma$
by cutting out $Im(g)$ (the new boundary component is an open string component $d$
parametrized outbound by arc length scaled to $1$).
The element $\Psi_t(x)$ of the right hand side of \rref{ek8} is now defined
by taking the vacuum element of $\Sigma_g$ (with $D$-brane choices induced by $c$),
and ``inserting'' (i.e. plugging in) the element $x\in\mathcal{K}_{aa}$ to $d$.
The main point of the operator $\Psi_t(x)$ is that its dependence on $g$ is
expressed by the formula
\beg{edef2}{\Psi_{t,gf}(x)=\Psi_{t,g}(x)\cdot f^{\prime}(0).
}
This follows from the fact that $x$ is a primary field of weight $1$.
But note that \rref{edef2} means that the element $\Psi_{t}(x)$ of the
right hand side of \rref{ek8} transforms as a $1$-dimensional measure, which
means precisely that it can be integrated without ``multiplying by $dt$''.
We must, of course, assume that the integral \rref{edef1} converges in the
subspace of trace class elements of the Hilbert tensor product.

\vspace{3mm}
Given that, however, \rref{edef1} can be considered an ``infinitesimal deformation''
in the following sense: Suppose we form the formal expression
\beg{edef3}{{U_{c^{\prime}}=U_{c} +\epsilon (\Delta U_c).}}
Then $U_{c^{\prime}}$ will satisfy gluing identities (=``Ward identities'')
up to linear terms in $\epsilon$. This is simply a consequence of additivity
of integration with respect to the integration locus.

\vspace{3mm}
It is important to note that $1$-dimensional anomaly doesn't spoil the
above construction. It is because, in the language of central extensions
of the worldsheet SLCMC, the line corresponding to $D^+$ (with the standard
parametrization of the open string component) is canonically isomorphic to $\C$,
and therefore the lines corresponding to the worldsheets $\Sigma$, $\Sigma^{\prime}$
are canonically isomorphic by gluing.

\vspace{3mm}

\section{Deformations of branes to perturbative level: breaking boundary
conformal invariance and the Kondo flow}

\label{sp}

Having an infinitesimal deformation does not guarantee that it can be
exponentiated to a finite deformation, or even to perturbative level
(i.e. to a Taylor series expansion in a formal deformation parameter $u$).
Let us look at the example of 
the Kondo effect in WZW branes. For this example, consider
the level $k$ WZW model for $SU(n)$. In the simplest case (sufficient for
our purposes), one considers the Cardy brane $a$ corresponding to the
$0$ label (we denote it by $a$ instead of $0$ to avoid confusion due to
overuse of that symbol). Now the primary fields we wish to consider are
in the state space $\mathcal{K}_{Va,Va}$ where $V$ is the target vector
space of an irreducible representation of $SU(n)$ which corresponds
to a label $\lambda\in P_k$.
We then have
\beg{eec1}{\mathcal{K}_{Va,Va}=Hom(V,V)\otimes \mh_{0}.
}
We shall sometimes write $\mathcal{K}_{aa}$ instead of $\mathcal{K}_{V_a,V_a}$
for brevity.
The primary fields in \rref{eec1} we wish to use for deformation
are of the form
\beg{eec2}{\phi_{\lambda}=\cform{\sum}{\alpha}{} S_{\alpha}J^{\alpha}_{-1}
}
where $\alpha$ runs through a basis of the Lie algebra $su(n)$, $S:su(n)\r
Hom(V,V)$ is the representation corresponding to $\lambda$ on the level
of Lie algebras, and $J^{\alpha}_{-1}$ is the weight $1$ primary field in 
$\mh_{0}=\mathcal{K}_{00}$ corresponding to $\alpha$ (realizing the
action of $su(n)$ on $\mh_{0}$ by vertex operators).

\vspace{3mm}
Let us compute some information about this infinitesimal deformation
explicitly. Specifically, we let us consider the infinitesimally deformed
operator $L_p(\epsilon)$ 
where $L_p$ corresponds to the vector field $z^{p-1}\partial/\partial z$
in the upper semicircle $S^{1+}=\{z\in \C|\;||z||=1,Im(z)\geq 0\}$.
But this deformation is easy to compute, since we must simply add two summands,
one of which is insertion of the vertex operator corresponding to \rref{eec2}
at $z=1$, and the other at $z=-1$; the second term must have the sign $(-1)^{p-1}$.
One thing to note is that we must be careful with the order of matrix
multiplicaton due to the fact that $Hom(V,V)$ is non-commutative. In fact,
neglecting terms of higher than linear order in $\epsilon$, we get
\beg{ebr1}{L_p(\epsilon)=\cform{\sum}{n\in\Z,\alpha}{}:J^{\alpha}_{n}J^{\alpha}_{-n+p}:
+\epsilon (S_{\alpha}^{L}J^{\alpha}_{n}+(-1)^{p+n+1}S_{\alpha}^{R}J^{\alpha}_{n})
}
where the superscripts $L,R$ indicate that the matrix in question should 
multiply the $Hom(V,V)$ factor from the left (resp. right).
As usual, $::$ denotes normal ordering. Another comment is that in the
first summand of \rref{ebr1}, one of the $\alpha$ indexes should be lowered.
However, it is customary to write the term in the present form. 
This is correct if the $\alpha$'s form an orthonormal basis.

\vspace{3mm}
Now consider the level one primary field $\phi_{\lambda}$ in $\mh_0$. 
Looking for a primary field of general weight $w$ in the infinitesimally deformed
sector $\mathcal{K}_{\alpha(\epsilon),\alpha(\epsilon)}$ of the form
\beg{ekon+}{\phi_{\lambda}(\epsilon)=\phi_{\lambda} +\epsilon\psi_{\lambda},
}
we get the equation
\beg{ekon++}{\begin{array}{lll}
L_p(\epsilon)\phi_{\lambda}(\epsilon) & =0 & \text{for $p>0$}\\
&=w\phi_{\lambda}(\epsilon)& \text{for $p=0$}
\end{array}
}
which must be satisfied up to (and including) linear order in $\epsilon$.
Expanding \rref{ekon++} using \rref{ekon+}, we see that we must have
$w=1+v\epsilon$ to get the $\epsilon$-constant terms to match, and
the $\epsilon$-linear terms give that
$$\cform{\sum}{n\in\Z,\alpha}{}:J^{\alpha}_{n}J^{\alpha}_{-n+p}:
\psi_{\lambda}+\cform{\sum}{n\in\Z,\alpha}{}(
S^{L}_{\alpha}J^{\alpha}_{n} + (-1)^{p+n+1}S^{R}_{\alpha}J^{\alpha}_{n})
\phi_{\lambda}
$$
is equal to $0$ for $p>0$ and to 
$$v\phi_{\lambda}+\psi_{\lambda}$$
for $p=0$. Now the condition for $p=0$ gives
\beg{ekon+++}{\cform{\sum}{n\in\Z,\alpha}{}:J^{\alpha}_{n} J^{\alpha}_{-n}:
\psi_{\lambda} +\cform{\sum}{n\in\Z,\alpha,\beta}{}(S^{L}_{\alpha}J^{\alpha}_{n}
+(-1)^{n+1}S^{R}_{\alpha}J^{\alpha}_{n})S_{\beta}J^{\beta}_{-1}=v\phi_{
\lambda}+\psi_{\lambda}.
}
Putting 
$$\psi_{\lambda}=\cform{\sum}{n>0}{}\psi_{\lambda}(n)$$
where $\psi_{\lambda}(n)$ is homogeneous of weight $n$, we get from 
\rref{ekon+++} the equation
\beg{ekoni}{\begin{array}{lll}
(n-1)\psi_{\lambda}(n)+\cform{\sum}{\alpha,\beta}{}
(S^{L}_{\alpha}J^{\alpha}_{n-1}+(-1)^{n}S^{R}_{\alpha}J^{\alpha}_{n-1})
S_{\beta} J^{\beta}_{-1}&=0&\text{if $n\neq 1$}\\
&=v\sum S_{\alpha}J^{\alpha}_{-1}& \text{if $n=1$.}
\end{array}
}
We see that \rref{ekoni} for $n\neq 1$ can be solved, yielding
\beg{ekonii}{\psi_{\lambda}(n)=\cform{\sum}{\alpha, \beta}{}
(S^{L}_{\alpha}J^{\alpha}_{n-1}+(-1)^n S^{R}_{\alpha})J^{\alpha}_{-1}/(1-n).
}
For $n=1$, however, we get 
\beg{enn*}{\cform{\sum}{\alpha,\beta}{}S_{[\alpha,\beta]}J^{[\alpha,\beta]}_{-1}=
v\sum S_{\alpha}J^{\alpha}_{-1}.}
This is a condition on $v$, which we easily see can force it to
be non-zero. For a precise evaluation of $v$, we must look more closely
at \rref{eec2}. To lower one of the $\alpha$ indices in that formula,
we are using the Killing form in the Lie algebra $g$ scaled
by some positive real factor. In other
words, if
$$\iota\in g\otimes_{\R} g$$
is the inverse of the scaled Killing form, then we have
$$\phi_{\lambda}= (S\otimes 1)\iota.$$
Therefore, $v$ depends inverse-linearly on the 
scaling factor of the Killing form.
For $g=su(n)$, we may as well complexify. We have 
$$su(n)_{\C}=sl(n,\C).$$
Upon complexification, the Killing form extends to a complex 
symmetric bilinear
form (not a Hermitian form). Let us normalize
by taking the symmetric bilinear form 
on $sl(n,\C)$
which is the restriction of the symmetric
bilinear form $B$ on $gl(n,\C)$ defined by
$$B(e_{ij},e_{k\ell})=-\delta_{i}^{\ell}\delta_{j}^{k}.$$
We are trying to calculate
$$\left(\begin{array}{c}
[\;,]\\ \otimes\\
\protect[\;,]\end{array}\right)
(B^{-1}\otimes B^{-1}).$$
We know this is invariant, hence equal to $vB^{-1}$ for some $v$
(see \rref{enn*}.
To calculate $v$, we can then select a basis element $x$
of $sl(n)$ with 
$$B(x,x)=1,$$
say, 
\beg{ekonbeta}{x=\frac{i}{\sqrt{2}}(e_{11}-e_{22}).
}
Then we have
\beg{ekongamma}{v=\cform{\sum}{i,j,k,\ell}{}B([e_{ij},e_{k\ell}],x)^2.
}
The non-zero contributions to \rref{ekongamma} are only when $\ell=j$, $k=i$,
and the contribution is $2$ when $\{i,j\}=\{1,2\}$ and $1/2$
when $|\{i,j\}\cap\{1,2\}|=1$ (and $0$ otherwise). Thus,
$$v=2\cdot(2+2(n-2)\cdot\frac{1}{2})=2n.$$
Now it is easily seen that the condition \rref{ekonii} indeed implies
\rref{ekon++} for all $p$, so we obtain a primary field $\phi_{\lambda}(\epsilon)$
of weight $1+\epsilon v$ to linear order in $\epsilon$. In other words,
a primary field of weight $1$ is not possible.

\vspace{3mm}
Before saying anything about higher powers of $\epsilon$, however, let
us ask what would be the effect of deforming boundary sectors
$\mathcal{K}_{aa}$ of CFT's defined on the SLCMC of closed/open worldsheets
along arbitrary fields. Just as in the last section, we have
an equation of the form
\beg{ekona}{\frac{dU_{\Sigma(\epsilon)}}{d\epsilon}=\int_c J(u).
}
Here $\Sigma(\epsilon)$ is the worldsheet $\Sigma$
with brane labels $a$ replaced by $a(\epsilon)$, 
$c$ is again the union of brane components labelled by $a(\epsilon)$.
$J(u)$ is again the element obtained by cutting out a holomorphic image,
under a function $f$,
of a standard half-disk $D^+$ from $\Sigma(\epsilon)$ such that the
brane component of $D^+$ lies on $c$ and is mapped isometrically.
(To be precise, near a boundary point of $c$, only a neighborhood of
$0$ in $D^+$ will lie on $c$, and the condition must be modified
accordingly, using analytic continuation.) Assuming all necessary convergence
works, we obtain a theory which ``breaks conformal invariance only
on $c$'', i.e. is a CFT defined on a modification of the SLCMC of 
worldsheets $\mathcal{D}$ where now each worldsheet is endowed with
an analytic metric on the brane components. 

\vspace{3mm}
One can attempt to write perturbative formulas for the deforming field
and calculate the deformed worldsheet vacua by continuing
the method just outlined. We have to realize however that there
are delicate convergence questions. For example, the field $\psi_{\lambda}$
is not an element of the Hilbert space $\mathcal{K}_{aa}$ (has ``infinite
norm''), so the norm would have to be also updated along with the 
deformation. Another issue (as we will see) is that the wrong choice
of the deformed field $\phi_{\lambda}$ (for example making it constant)
will lead to non-convergence even in a more profound sense, i.e. in  
the product of the weight-homogeneous summands of $\mathcal{K}_{aa}$.
One therefore needs a method for keeping track, at least to some degree,
of the convergence properties of the deformed vacua. 

\vspace{3mm}
Before starting this discussion, it is helpful to realize that there is 
at least one simplification in the boundary sector deformation questions
which is not available in the case of bulk deformations (which we
do not discuss here). Namely, to characterize vacua of all worldsheets
whose brane components, in the alias notation, are labelled by
known branes and $a(\epsilon)$, it suffices to know the vacuum 
$u_{\epsilon}(t)$ in (some completion of) $\mathcal{K}_{aa}$ corresponding
to the worldsheet $D^{+}_{a(\epsilon),t}$, which is the standard half-disk
$D^+$ with brane component labelled by $a(\epsilon)$ and metric which is
scaled length in such a way that the total length of the brane component
is $2t$. The reason this suffices is that every worldsheet $B(\epsilon)$ 
whose brane components are labelled by known branes and $a(\epsilon)$
can be
glued from copies of $D^{+}_{a(\epsilon),t}$ and a worldsheet $\Sigma(\epsilon)$ which
is labelled by known branes and $a(\epsilon)$ and has the property that
all branes labelled by $a(\epsilon)$ have length $0$ (are single points).
But the vacuum associated with $\Sigma(\epsilon)$ should not be affected by
the deformation, i.e. we should have
$$U_{\Sigma(\epsilon)}=U_{\Sigma(0)}.$$
Therefore, $U_{B}$ can be calculated from $U_{\Sigma(0)}$ and $U_{D^{+}_{a(\epsilon),t}}=
u_{\epsilon}(t)$.
Now the worldsheets $\Sigma(\epsilon)$ is not really an element of our moduli
space, so the vacua $U_{\Sigma(\epsilon)}$ may really not be well defined
as trace-class elements of the corresponding Hilbert spaces (additional
completion may be required). Nevertheless, $U_{B}$ should still be
determined by limit arguments. We will not treat this in detail in this paper.

\vspace{3mm}
Nevertheless, the idea one gets from this is to rewrite the deformation
process in such a way that it refers only to the elements $u_{\epsilon}(t)$.
As it turns out, it is actually more advantageous to work with the
element
\beg{ekon1}{v_{\epsilon}(t)=A^{+}_{t}u_{\epsilon}(t)
}
where $A^{+}_{t}$ is the worldsheet obtained by cutting $tD^{+}_{a}$ out of $D^{+}_{a}$.
For brevity, we use the same notation for the associated operator $\mathcal{K}_{aa}\r
\mathcal{K}_{aa}$ (which would be more precisely denoted by $U_{A^{+}_{t}}$,
which however is ambigous, since it can be understood as an operator or
an element of $\mathcal{K}_{aa}^{*}\hat{\otimes} \mathcal{K}_{aa}$). Note that
in the operator $A^{+}_{t}$, we use the operator corresponding to the brane
$a$, not the deformed brane $a(\epsilon)$. Because of this discrepancy, note
that we will need to discuss convergence: let $K(n)\subset \mathcal{K}_{aa}$
be the subspace of elements of weight $n$. Then we have 
$$K=\cform{\bigoplus}{n\geq 0}{}K(n)\subset \mathcal{K}_{aa}\subset
\hat{K}=\cform{\prod}{n\geq 0}{}K(n).$$
Operators defined on $K$ may not extend to $\hat{K}$.
Note that by \rref{ekon1},
$v_{\epsilon}(t)$ is obtained from $u_{\epsilon}(t)$
by multiplying the weight $n$ summand by $t^n$. Our approach here
will be the same as in Remark 2 of Section \ref{s2a}. For the remainder
of this section, then, the elements discussed will be elements of
$\hat{K}$, and convergence of operators applied to each individual
element will be treated by means of analysis. In this setting, we will
construct rigorously the elements
$$v_{\epsilon}(t)\in\hat{K}[[\epsilon]],$$
i.e. the element $v_{\epsilon}(t)$ is an actual function of
$t$, but is perturbatively expanded in $\epsilon$.

\vspace{3mm}
Now the Kondo flow, as reviewed, say, in \cite{moore}, (note however
that in the present paper we do not consider supersymmetry), one expects the
equation
\beg{ekon2}{\frac{dv_{\epsilon}(t)}{dt}=\epsilon(Y(\phi_{\lambda},t)_L
+Y(\phi_{\lambda},-t)_{R})v_{\epsilon}(t)
}
(as before, the susbcripts $L$ and $R$ refer to whether the $Hom(V,V)$-matrix
from $\phi_{\lambda}$ multiplies the one from $v_{\epsilon}(t)$
from the left or right). The equation \rref{ekon2} is a direct analog
of the equation (5.30) from \cite{moore} (which refers to the unit disk with
a closed brane boundary component labelled by $a(\epsilon)$, while
\rref{ekon2} refers to $D^{+}_{a(\epsilon),t}$). From out point of view,
this corresponds to exponentiating the infintesimal deformations by
the {\em constant} field $\phi_{\lambda}(\epsilon)=\phi_{\lambda}$.
We shall discuss this point in more detail shortly.

\vspace{3mm}
Of course, we know $\phi_{\lambda}$ and its vertex operator, so
\rref{ekon2} becomes
\beg{ekon3}{\frac{dv_{\epsilon}(t)}{dt}=
\epsilon\cform{\sum}{n\in\Z}{}(S_{\alpha}^{L}-(-1)^{n-1}S_{\alpha}^{R})
J_{-n}^{\alpha}v_{\epsilon}(t)t^{n-1}.
}
Instead of writing the solution to \rref{ekon3} formally using the
path-ordered exponentiation $P\exp$, let us try to expand the
solution perturbatively in $\epsilon$. Setting
\beg{ekon4}{v_{\epsilon}(t)=\cform{\sum}{n\geq 0}{}v_{n}(t)\epsilon^n,
}
we get from \rref{ekon4}
\beg{ekon5}{\frac{dv_{\epsilon}(t)}{dt}=
\cform{\sum}{n\in\Z}{}(S_{\alpha}^{L}-(-1)^{n-1}S_{\alpha}^{R})
J_{-n}^{\alpha}v_{n-1}(t)\cdot t^{n-1}.
}
Assuming
\beg{ekon5a}{v_0(t)=1,
}
the equation for $v_1(t)$ from \rref{ekon5} works out well. We get
\beg{ekon6}{v_1(t)=2\cform{\sum}{n\geq 0}{}S_{\alpha}J^{\alpha}_{-2n-1}t^{2n+1}/
(2n+1)
}
which agrees with what we already know. The problem is, however, that
the equation \rref{ekon5} for $v_2$ diverges. Therefore, we are faced with
the problem of regularization.

\vspace{3mm}
We proceed as follows: first note that 
the equation \rref{ekon3} takes the infinitesimal
approach, composing $v_{\epsilon}(t)$ with ``$v_{\epsilon}(1/\infty)$'',
or more precisely with
\beg{ekon7}{\left.\frac{\partial v_{\epsilon}(t)}{\partial t}\right|_{t=0^+}.
}
This should be $2\phi_{\lambda}\epsilon$.
In fact, the idea is that the derivative of $v_{\epsilon}(t)$ at
$t=0$ should have the meaning of a worldsheet with ``infinitesimal''
$a(\epsilon)$-brane component. Therefore, it should be deformed
by simply integrating the field $\phi_{\lambda}(\epsilon)$, so in general,
\rref{ekon7} should be equal to
$$2\int_{0}^{\epsilon}\phi_{\lambda}(x)dx.$$
When $\phi_{\lambda}(\epsilon)=\phi_{\lambda}$ is constant, we get
$2\phi_{\lambda}\cdot\epsilon$. Composition of $D^{+}_{a(\epsilon),t}$
with the ``infinitesimal worldsheet'' $D^{+}_{a(\epsilon),dt}$ then
gives the equation \rref{ekon2}.
But we saw that assuming this leads to a divergent formula for
$v_{\epsilon}(t)$. Therefore, if the formula is to be regularized,
we must assume that $\phi_{\lambda}(\epsilon)$ cannot be constant
in $\epsilon$ and that moreover its coefficients at $\epsilon^1$
(which we formerly denoted by $\psi_{\lambda}$) must be divergent.
In other words, to regularize rigorously, we must consider the case
when the limit \rref{ekon7} does not exist at all.
To proceed,
we must replace \rref{ekon1} by an equation involving the finite worldsheet
$B_{s,t,a(\epsilon)}$ obtained from $D^{+}_{a(\epsilon),s+t}$
by cutting out $-sD^{+}_{a(\epsilon),t}$ and $tD^{+}_{a(\epsilon),s}$.
Identifying, again, notationally $B_{s,t,a(\epsilon)}$ with the corresponding
operator
$$K\otimes K\r \hat{K},$$
we have
$$B_{s,t,a}(u_{s}(\epsilon),u_{t}(\epsilon))=u_{s+t}(\epsilon),$$
which, in terms of vertex operators in $K$, can be written as
\beg{ekon8}{\exp(-sL_{-1})Y(v_{s}(\epsilon),s+t)v_{t}(\epsilon)=v_{s+t}(\epsilon).
}
(Recall again that this operator is independent of brane deformation because
the $a(\epsilon)$-brane component is degenerate to a discrete points;
in other words, the operators corresponding to $B_{s,t,a}$, $B_{s,t,a(\epsilon)}$
coincide.)
Let us attempt to solve \rref{ekon8} by perturbative expansion in $\epsilon$,
i.e. by setting \rref{ekon4}, and assuming \rref{ekon5a}, \rref{ekon6}.
(Actually, one has to verify that \rref{ekon6} is consistent, i.e.
makes \rref{ekon8} satisfied to linear order in $\epsilon$, but that is
easily done.) In order to make this work, we need an inductive convergence
assumption. Recall that for an element $u\in \hat{K}$, we denote by $u(n)$
its homogeneous summand of weight $n$. Our assumption is
\beg{ekon9}{
\parbox{3.5in}{For all $m$, $v_n(t)(m)$ is a $Hom(V,V)$-matrix of polynomials
of degree $\leq n$ in $\mh_{0}$ and $\partial_{t}^{k}$ of all its coefficients
are of order $o((t^{\prime})^m)$ for every $t^{\prime}>t$, $k\geq 0$.
}
}
By way of explanation, recall that $\mathcal{H}_{0}$ is a Hilbert
completion of an algebra with generators
$J_{-n}^{\alpha}$. The algebra is not commutative, but by a polynomial
of degree $\leq n$ we mean simply a sum of finitely many elements, 
each of which is a product of $\leq n$ of the generators. 

\vspace{3mm}
We emphasize that \rref{ekon9} is an induction hypothesis, i.e.
a statement which will be completely proved in the course of the 
construction. For $n=0,1$, the statement is trivial 
(see \rref{ekon5a}, \rref{ekon6}). The induction step, i.e.
the fact that assuming \rref{ekon9} with $n$ replaced by $n-1$
in fact implies the statement \rref{ekon9} for a given $n$,
will be given at the end of the construction.

\vspace{3mm}
Now assume $v_{1}(t),...,v_{n-1}(t)$ have been constructed and \rref{ekon9}
is satisfied with $n$ replaced by each of the numbers $1,...,n-1$.
Then the coefficient at $\epsilon^n$ of \rref{ekon8} gives the equation
\beg{ekon10}{\exp(-sL_{-1})v_n(t)+
\exp(tL_{-1})v_{n}(s)=v_{n}(s+t)+C(s,t).
}
The element $C(s,t)$ is known and converges by \rref{ekon9}, and further
satisfies
\beg{ekon11}{
\parbox{3.5in}{All the $C(s,t)(m)$'s are $Hom(V,V)$-matrices
of polynomials in $\mathcal{H}_{0}$ of degree $\leq n$ and all
$\partial_{t}^{k}\partial_{s}^{\ell}$'s of all their coefficients
are $o((t^{\prime})^m$ for every $t^{\prime}>s+t$, $k,\ell\geq 0$.}
}
The functional equation \rref{ekon10} is solved as follows: first notice
that
\beg{ekon12}{(\frac{\partial}{\partial s}+L_{-1})\exp(-sL_{-1})v_{n}(t)=0,
}
\beg{ekon13}{(\frac{\partial}{\partial t}-L_{-1})\exp(tL_{-1})v_{n}(s)=0.
}
Further, the operators 
$$\frac{\partial}{\partial s}+L_{-1},\;\frac{\partial}{\partial t}-L_{-1}$$
commute, so we get from \rref{ekon12}, \rref{ekon13}
\beg{ekon14}{(\frac{\partial}{\partial s}+L_{-1})(\frac{\partial}{\partial t}-L_{-1})
v_{n}(s+t)=-
(\frac{\partial}{\partial s}+L_{-1})(\frac{\partial}{\partial t}-L_{-1})C(s,t).
}
But
$$\frac{\partial}{\partial s}f(s+t)=\frac{\partial}{\partial t}f(s+t)=
f^{\prime}(s+t),$$
so from \rref{ekon14},
$$-(\frac{\partial}{\partial s}+L_{-1})(\frac{\partial}{\partial t}-L_{-1})C(s,t)
=h(s+t)$$
for some function $h$. If we denote, for the moment, ordinary
differentiation by $D$, \rref{ekon14} also gives
\beg{ekon15}{(D+L_{-1})(D-L_{-1})v(s)=h(s)
}
(where $v(s)=v_n(s)$), or
\beg{ekon16}{v^{\prime\prime}(s)-L^{2}_{-1}v(s)=h(s).
} 
The basis of solutions of the homogeneous equation corresponding to
\rref{ekon16} are the solutions to the equations
\beg{ekon17}{v^{\prime}(s)=L_{-1}v(s),
}
\beg{ekon17a}{v^{\prime}(s)=-L_{-1}v(s),
}
which would be
\beg{ekon18}{v(s)=\exp(L_{-1}s)u,
}
\beg{ekon19}{v(s)=\exp(-L_{-1}s)u
}
for an arbitrary field $u$, respectively, if $L_{-1}$
were invertible. In fact, note that 
\beg{ekon20}{v(s)=\frac{\exp(L_{-1}s)-\exp(-L_{-1}s)}{2L_{-1}}u
}
spans the space of solutions to the homogeneous functional equation
corresponding to \rref{ekon10}. Thus, the actual basis of solutions 
consists of one of the functions \rref{ekon18}, \rref{ekon19},
and the function \rref{ekon20}. We may now explicitly solve
the non-homogeneous differential equation \rref{ekon16} by
variation of constants: if 
$$Y^{\prime}(s)=A(s)Y(s)$$
is a basis of solutions for a homogeneous first order system, then
$$Z^{\prime}(s)=A(s)Z(s)+K(s)$$
is solved by
\beg{ekon20a}{Z(s)=Y(s)\int Y(s)^{-1}K(s)ds.
}
In any case, we know how to solve the non-homogeneous differential
equation \rref{ekon15}. To solve the non-homogeneous functional equation
\rref{ekon10}, we first must check that a solution of \rref{ekon10}
exists. For that, there is a necessary and sufficient {\em cocycle 
condition} on $C(s,t)$:
\beg{ekon21}{\exp(uL_{-1})C(s,t)+C(s+t,u)=C(s,t+u)+\exp(-sL_{-1})C(t,u).
}
There is a standard reason why the cocycle condition is always satisfied
in this situation: Consider the element
$$y(s)=\cform{\sum}{m=0}{n-1}v_{m}(s)\epsilon^m.$$
Now consider
$$D(s,t)=B_{s,t,a}^{+}(y(s),y(t))-y(s+t).$$
Then $D(s,t)$ satisfies a cocycle condition coming from the associativity
of composition:
\beg{ekon22}{B_{s+t,u,a}^{+}(B^{+}_{s,
t,a}(u_1,u_2),u_3)=B^{+}_{s,t+u,a}(u_1,B^{+}_{t,u,a}(u_2,u_3)).
}
But the $\epsilon^n$-coefficient of $D(s,t)$ is $C(s,t)$, the
$\epsilon^n$-coefficient of \rref{ekon22} is \rref{ekon21}.

\vspace{3mm}
Therefore, the functional equation \rref{ekon10} has a solution $v_n$.
If $v$ is a solution of the differential equation \rref{ekon15},
then $v_n-v$ is a solution of the corresponding homogeneous differential
equation. But as we saw, we know all such solutions, and we further
saw that the solutions \rref{ekon20} satisfy the homogeneous functional
equation corresponding to \rref{ekon10}. Therefore, to solve \rref{ekon10},
plug in a general solution \rref{ekon20a} of \rref{ekon15} to \rref{ekon10}
and calculate (up to the indeterminacy \rref{ekon20}) the constants
from the resulting system of consistent (underdetermined) linear equations.

\vspace{3mm}
{\bf The induction step:}
There remains the subtle question whether the condition \rref{ekon9} is
satisfied for any of the solutions $v_n(t)$ just constructed. This can
be shown as follows. First, call any function of $t>0$ satisfying
\rref{ekon9} {\em properly bounded}. We shall prove that several functions
have this property. First, we know from \rref{ekon11} that
\beg{ekon23}{\text{$(D^2-L_{-1}^{2})v_n(t)$ is properly bounded}
}
for any solution $v_{n}(t)$ of \rref{ekon10}. Next, consider
\beg{ekon24}{(D+L_{-1})v_n(t)=:w(t).
}
Applying 
$$\frac{\partial}{\partial s} +L_{-1}$$
to \rref{ekon10} (or alternately, exponentiating \rref{ekon23}),
we see that
\beg{ekon25}{\text{$\exp(tL_{-1})w(s)-w(s+t)$ is properly bounded.}
}
It follows from \rref{ekon25} that if $w(s_0)$ is properly bounded,
then so is $w(s)$ for all $s>s_0$. Unfortunately, that is not enough.
However, we can remedy the situation by using the following trick:
$w(t)$ is a solution to the first order system of differential equations
\beg{ekon26}{(D-L_{-1})w(t)=h(t).
}
But \rref{ekon26} is partially decoupled with respect to conformal
weight: For each weight $n$, we have an equation of the form
$$Dw(t)(n)=
L_{-1}w(t)(n-1)+\phi(t)(n).$$
This means that we can apply initial conditions successively
at weights $0,1,2,...$ at points $t_{0}>t_{1}>t_{2}>...$ where
$t_{n}\r 0$. It follows from the definition and \rref{ekon25}
that choosing the initial conditions
\beg{ekon27}{ w(t_n)(n)=0}
implies that
\beg{ekon28}{\text{$w(t)$ is properly bounded for all $t>0$.}
}
(Note: plugging in \rref{ekon20} for $v_{n}(t)$ gives
$w_{homog}(t)=\exp(tL_{-1})u$, thus showing that the initial conditions
for $w$ can indeed be chosen arbitrarily for a solution $v_{n}(t)$
of the non-homogeneous functional equation \rref{ekon10}.)
Now suppose $v_{n}(t_{0})$ is not properly bounded where $v_{n}(t)$
solves \rref{ekon10} and \rref{ekon24}. Then by \rref{ekon24},
\beg{ekon29}{
\parbox{3.5in}{$\exp(-s(D+L_{-1}))v_{n}(t)=\exp(-sL_{-1})v_{n}(t_0-s)-
v_n(t_0)$ is properly bounded for all $0<s<t_0$}
}
which means
\beg{ekon30}{
\parbox{3.5in}{$v_n(t)$ is not properly bounded for any $0<t<t_0$.}
}
Choosing $0<s,t$ such that $s+t<t_0$, \rref{ekon10} can be rewritten
as
\beg{ekon31}{\exp(tL_{-1})v_n(s)=-\exp(-s(D+L_{-1}))v_n(s+t)+C(s,t).
}
The right hand side is properly bounded by \rref{ekon24}, \rref{ekon28},
while the left hand side is not by \rref{ekon30} - a contradiction.

(In more detail, the first term of the right hand side of \rref{ekon31}
is 
\beg{ennn*}{\exp(-sL_{-1})v_{n}(t)-v_{n}(s+t).
}
Treating $r=s+t$ as a constant, \rref{ennn*} is
\beg{ennn+}{\exp(-sL_{-1})v_{n}(r-s)-v_{n}(r).
}
Applying $\frac{\partial}{\partial s}$, we get
$$\exp(-sL_{-1})(-L_{-1}-\frac{\partial}{\partial s})v_{n}(r-s)$$
which is properly bounded for all $s$ by \rref{ekon28}. Now integrate
from $0$ to $s$ by $ds$.)

\vspace{3mm}
Thus, we have defined a process which constructs the vacua $u_{\epsilon}(t)$
expanded in the perturbative parameter $\epsilon$. The process is
convergent in the sense of QFT. We also saw that the process has infinitely
many degrees of freedom due to the indeterminacy in solving the
equation \rref{ekon10} at every step (power of $\epsilon$). It is possible
that some natural normalization of this process exists, but it cannot
be obtained by considering the limit \rref{ekon7}, which does not
exist. It is possible that the limit brane $\mathcal{H}_{\lambda}$
can be obtained as a non-perturbative limit of this process, i.e.
as an analytic continuation of one of the theories constructed, at
a particular value of $\epsilon$. To that end, however, several steps
are still missing. Let us recapitulate what we constructed. The vacua
$u_{\epsilon}(t)$ which we did construct determine all vacua corresponding
to worldsheets $X_{\epsilon}$ with brane components labelled by $a(\epsilon)$ or
known (Cardy) branes. Further, the elements $u_{\epsilon}(t)$ converge
as formal series in the parameter $\epsilon$. This is what we mean 
by convergence in the sense of QFT. What we haven't shown is that
the vacua corresponding to all the worldsheets $X_{\epsilon}$
converge and in what sense. To do so, an inner product dependent on $\epsilon$
would have to be constructed on $K$. We would then define 
$\mathcal{K}_{a(\epsilon),a(\epsilon)}$ as the Hilbert completion
of $K$ under this inner product, and then we would have to show that
our recipe for calculating vacua indeed gives trace class elements in
the appropriate Hilbert tensor products. All this would be needed to
actually prove that we have constructed a D-brane category over worldsheets
breaking conformal invariance on brane components labelled by $a(\epsilon)$.
Nevertheless, all these statements, which we do not prove here, make
plausible conjectures.

\vspace{3mm}
In the next section, we describe a formalism 
of ``topological $D$-brane categories'', which captures
such conjectured process. At the end of the next section,
we shall state the Kondo effect conjecture (or more precisely one of its
consequences) in a more precise way.

\vspace{5mm}

\section{A general mathematical formalism for describing Kondo-like effects}

\label{s4}

The point of this section is mostly to introduce a {\em topological} version
of the concepts introduced in Section \rref{s2}. 

\vspace{3mm}
First, we let the topology on $\C_2$, the category of finite-dimensional
$\C$-vector spaces, be discrete on objects, and the standard topology on morphisms
(induced by the product topology of copies of $\C$).

\vspace{3mm}
Now the $D$-brane space is a $2$-vector space $\mathcal{B}$, which is
free on a set $B$, but we can
no longer assume that it is finite-dimensional (i.e. that $B$ is
finite), as we clearly need infinitely
many non-isomorphic branes to reproduce a Kondo-like effect. Now
we further assume that $\mathcal{B}$ is {\em topologized}. By this we
will mean that $\mathcal{B}$ has a topology on its sets of objects,
and morphisms, which makes all the categorical operations
continuous, and also the $\C_2$-module structure continuous. Note that
(and this is important) we do not assume that $B=B\cdot\C$ would be
a closed subset of $\mathcal{B}$, i.e. we allow (as in the Kondo-effect)
that ``elementary'' branes (by which we mean branes in $B$) converge to 
linear combinations of elementary branes.

\vspace{3mm}
The first thing to notice is that since $\mathcal{B}$ is infinite-dimensional,
we don't have an equivalence
$\mathcal{B}^{**}\simeq \mathcal{B}$
or \rref{eoc10}. We will, therefore, as usual instead use the embedding
\beg{eoc11}{\mathcal{B}\r\mathcal{B}^{*}
}
using the basis $B$ of ``elementary branes'' (the map \rref{eoc11} is
given by sending $b\in B$ to the map which assigns to a linear combination
its coefficient at $b$). Now clearly introducing the map \rref{eoc11}
breaks (some) functoriality, and hence, it will also break continuity,
since we are not assuming that $B$ is a closed set in $Obj(\mathcal{B})$.

\vspace{3mm}
Next, we must then discuss the open string sector $\mathcal{K}$. According
to the convention just described, we will simply specify, for each pair
of elementary branes $a,b\in B$, a Hilbert space $\mathcal{K}_{ab}$.
We can extend this to a map of $\C_2$-modules 
\beg{econt1}{\mathcal{K}:\mathcal{B}\otimes\mathcal{B}\r \C_{2}^{Hilb}}
in the canonical way. It is, however, too strong to assume that the functor
\rref{econt1} is continuous on the nose, since we have a discrete topology
on the target. A similar issue arose in \cite{hk} when we proposed a topological
group completion of $\C_2$. Here, however, we shall deal with it in a more
straightforward way. We simply require that 
for each object $a\in Obj(
\mathcal{B})$, we are given an open neighborhood $U_a$ of $a$ in $Obj(\mathcal{B})$,
and for each $x\in U_a$, $y\in U_b$
an isomorphism of vector spaces
\beg{econt2}{\phi_{xy}^{ab}:\mathcal{K}_{ab}\r\mathcal{K}_{xy}.}
Furthermore, when $z\in U_x\cap U_a$, $t\in U_y\cap U_b$, we have
\beg{econt3}{\phi_{zt}^{ab}=\phi_{zt}^{xy}\phi_{xy}^{ab}.}
The morphisms $\phi_{xy}^{ab}$ are also required to be preserved by addition and
scalar multiplication in $\mathcal{B}$, and to form commutative diagrams with
coherence isomorphisms of those operations, as usual.

\vspace{3mm}
We want to introduce the SLCMC $\mathcal{C}(\mathcal{B},\mathcal{K},\mathcal{H})$
in this topological case, and also its $1$-dimensional anomaly version
$\tilde{\mathcal{C}}(\mathcal{B},\mathcal{K},\mathcal{H})$.
$\mathcal{H}$ is
as usual, $\mathcal{B}, \mathcal{K}$ are as above.
We shall only explain what changes in the definition of
sections of
$\mathcal{C}(\mathcal{B},\mathcal{K},\mathcal{H})$
over a point, since the other points are routine.  A section over a point
over a graph $\Gamma$ with $p$ closed components, $q$ open pure $D$-brane
components and $s$ open components with $\ell_j$ open brane components, $j=1,...,s$,
specifies for each choice $c$ of 
elementary branes $a_{jm}\in B$, where 
$m\in\{1,...,\ell_j\}$, (as usual, we put $a_{j0}=a_{jm}$) and elementary
branes $a_i\in B$ for $i=1,...q$ an element
\beg{evac1}{U_c\in \cform{\bigotimes}{i=1}{p} \mathcal{H}^{(*)}\otimes
\cform{\bigotimes}{j=1}{s}\left(
\cform{\bigotimes}{m=0}{\ell_j-1} \mathcal{K}_{a_{jm},a_{j,m+1}}^{(*)}
\right),
}
subject to certain conditions
(again, completions of tensor products
by taking Hilbert tensor product and then specializing to
trace class elements are suppressed from the notation, and ${}^{(*)}$ means
that Hilbert dual is taken on inbound string components, as is Section 
\rref{s2} above).

\vspace{3mm}
The condition we need on \rref{evac1} is continuity. To formulate that,
we first note that we can extend the choices $c$ allowed in \rref{evac1}
to $a_{jm}, a_i\in Obj(\mathcal{B})$ as follows: each time we multiply an elementary
brane corresponding to
an open brane component by a vector space $V$, the corresponding target space of 
\rref{evac1} is multiplied by $V\otimes V^*$ because two open string components
end on
that brane component. Simply then multiply the vacuum vector $U_c$ by the identity
in $V\otimes V^*$. An analogous prescription works also for forming sums, which
allows us to plug in any linear combinations of elementary branes.
For consistency reasons, closed brane components also enter the picture:
when multiplying a brane $a_i$, $i=1,...,q$, by $V$, the vacuum vector is
multiplied by $dim(V)$ (and is also additive on taking sums of branes in the
place of $a_i$).

\vspace{3mm}
Now the continuity condition on vacua simply states that for 
any convergent net $x_{jm}\r a_{jm}$,
$x_{i}\r a_i$, the vacua $U_{c_x}$ converge to $U_{c_a}$ using, where necessary,
the transformations $\phi$ of \rref{econt2} to identify target spaces.

\vspace{3mm}
We may now state formally the conjecture about the WZW model. Unfortunately,
the formulation still is not completely satisfactory, due to the fact that
as commented in \cite{hk}, we still have no mathematical formalism for 
asking what {\em are} the branes of a closed CFT: even if we add a
brane $a$, and specify
the sector of open strings ending on that brane $a$, 
it will not explain the sectors of strings beginning on $a$ and ending on
another brane. From a physical point of view, this is an example of {\em
locality violation} (see comments at the end of section 5.5. in \cite{moore}).

\vspace{3mm}
Remarkably, however, we saw in Section \ref{s3} that at least infinitesimally,
the physical recipe for
{\em deformations} to a brane update automatically all sectors of strings whose
one or both ends lie on the deformed brane, and that local deformations
(on CFT's with source $\mathcal{S}_b$) correspond
to weight $1$ fields. Further, this effect is maintained by
exponentiation, at least to perturbative level. Therefore, we can state our
conjecture as follows:

\vspace{3mm}
\noindent
{\bf Conjecture:} Let $\mathcal{H}$ be the closed sector of
the level $k$ WZW model for $SU(n)$. Then there exists an SLCMC
$\mathcal{C}(\mathcal{B}, \mathcal{K},\mathcal{H})$ and
a CFT 
$$\mathcal{S}_b\r \mathcal{C}(\mathcal{B}, \mathcal{K},\mathcal{H})
$$
with the following
properties:

\vspace{3mm}
\begin{enumerate}
\item
Every object of $\mathcal{B}$ is (isomorphic to an object) in the path-component
of a (finite vector space-valued) linear combination of Cardy branes.

\vspace{3mm}
\item
The neighborhoods $U_a$ for $a\in Obj{\mathcal{B}}$ are homeomorphic to
the (real parts of) vector spaces of weight $1$ vectors in $\mathcal{K}_{aa}$,
such that for $x\in U_a$, the transition functions between $U_a\cap U_x$ and
$U_a$ are smooth real-analytic (with a convergent exponentiation map
described by the procedure in the last section).

\vspace{3mm}
\item
The set of (path)-connected components of $\mathcal{B}$ is isomorphic to
$$\Z/((k+n)/gcd(k+n, lcm(1,...,n-1)))$$ 
via the map which sends $a$ to the dimension
of the bottom conformal weight 
space of the corresponding irreducible level $k$ representation
of $\tilde{L}SU(n)$. 

\end{enumerate}

\end{document}